\documentstyle[graphicx,ulem,amsmath]{mn2e}

\begin{document}

\title[GAMA Survey: RSD with multiple tracers]{Galaxy And Mass
  Assembly (GAMA): improved cosmic growth measurements using multiple
  tracers of large-scale structure}

\author[Blake et al.]{\parbox[t]{\textwidth}{Chris
    Blake$^1$\footnotemark, I.K.Baldry$^2$, J.Bland-Hawthorn$^3$,
    L.Christodoulou$^4$, M.Colless$^5$, \\ C.Conselice$^6$,
    S.P.Driver$^{7,8}$, A.M.Hopkins$^9$, J.Liske$^{10}$,
    J.Loveday$^{11}$, P.Norberg$^{12}$, \\ J.A.Peacock$^{13}$,
    G.B.Poole$^{14}$, A.S.G.Robotham$^{7,8}$ } \\ \\ $^1$ Centre for
  Astrophysics \& Supercomputing, Swinburne University of Technology,
  P.O. Box 218, Hawthorn, VIC 3122, Australia \\ $^2$ Astrophysics
  Research Institute, Liverpool John Moores University, IC2, Liverpool
  Science Park, 146 Brownlow Hill, Liverpool, L3 5RF, U.K. \\ $^3$
  Sydney Institute for Astronomy, School of Physics, University of
  Sydney, NSW 2006, Australia \\ $^4$ Institute of Cosmology \&
  Gravitation, Dennis Sciama Building, University of Portsmouth,
  Portsmouth, PO1 3FX, U.K. \\ $^5$ Research School of Astronomy and
  Astrophysics, The Australian National University, Canberra, ACT
  2611, Australia \\ $^6$ School of Physics \& Astronomy, University
  of Nottingham, University Park, Nottingham, NG7 2RD, U.K. \\ $^7$
  International Centre for Radio Astronomy Research (ICRAR),
  University of Western Australia, Crawley, WA 6009, Australia \\ $^8$
  SUPA, School of Physics and Astronomy, University of St Andrews,
  North Haugh, St Andrews, KY16 9SS, U.K.  \\ $^9$ Australian
  Astronomical Observatory, P.O.~Box 915, North Ryde, NSW 1670,
  Australia \\ $^{10}$ European Southern Observatory,
  Karl-Schwarzschild-Str.~2, 85748 Garching, Germany \\ $^{11}$
  Astronomy Centre, University of Sussex, Falmer, Brighton BN1 9QH,
  U.K. \\ $^{12}$ Institute for Computational Cosmology, Department of
  Physics, Durham University, South Road, Durham DH1 3LE,
  U.K. \\ $^{13}$ Institute for Astronomy, University of Edinburgh,
  Royal Observatory, Blackford Hill, Edinburgh EH9 3HJ, U.K.
  \\ $^{14}$ School of Physics, University of Melbourne, Parkville,
  VIC 3010, Australia}

\maketitle

\begin{abstract}
  We present the first application of a `multiple-tracer'
  redshift-space distortion (RSD) analysis to an observational galaxy
  sample, using data from the Galaxy and Mass Assembly survey (GAMA).
  Our dataset is an $r < 19.8$ magnitude-limited sample of $178{,}579$
  galaxies covering redshift interval $z < 0.5$ and area $180$
  deg$^2$.  We obtain improvements of $10$-$20\%$ in measurements of
  the gravitational growth rate compared to a single-tracer analysis,
  deriving from the correlated sample variance imprinted in the
  distributions of the overlapping galaxy populations.  We present new
  expressions for the covariances between the auto-power and
  cross-power spectra of galaxy samples that are valid for a general
  survey selection function and weighting scheme.  We find no evidence
  for a systematic dependence of the measured growth rate on the
  galaxy tracer used, justifying the RSD modelling assumptions, and
  validate our results using mock catalogues from N-body simulations.
  For multiple tracers selected by galaxy colour, we measure
  normalized growth rates in two independent redshift bins $f
  \sigma_8(z=0.18) = 0.36 \pm 0.09$ and $f \sigma_8(z=0.38) = 0.44 \pm
  0.06$, in agreement with standard GR gravity and other galaxy
  surveys at similar redshifts.
\end{abstract}

\section{Introduction}
\renewcommand{\thefootnote}{\fnsymbol{footnote}}
\setcounter{footnote}{1}
\footnotetext{E-mail: cblake@astro.swin.edu.au}

The large-scale structure of the Universe is one of the most valuable
probes of the cosmological model, enabling measurements to be
performed of the cosmic distance-scale and expansion rate, the
constituents of the Universe, and the gravitational forces which drive
the growth of structure with time.  In particular, the `gravitational
growth rate' is accessible through the imprint of redshift-space
distortion (RSD) in the pattern of structure.  RSD describes the
apparent anisotropic clustering induced by the small shifts in galaxy
redshifts that result from the correlated peculiar velocities that
galaxies possess in addition to the underlying Hubble-flow expansion.

This cosmic structure has been mapped out by a sequence of galaxy
redshift surveys such as the 2-degree Field Galaxy Redshift Survey
(2dFGRS, Colless et al.\ 2001), the 6-degree Field Galaxy Survey
(6dFGS, Jones et al.\ 2009), the Sloan Digital Sky Survey (SDSS, York
et al.\ 2000), the WiggleZ Dark Energy Survey (Drinkwater et
al.\ 2010) and the Baryon Oscillation Spectroscopic Survey (BOSS,
Dawson et al.\ 2013).  The accuracy of cosmological measurements are
often limited by `sample variance', the inherent fluctuations between
different portions of the Universe.  In order to obtain more precise
measurements, the scientific progress of galaxy redshift surveys has
emphasized mapping ever-greater cosmic volumes, often targetting a
relatively sparse distribution of a single type of galaxies (with
number density $\sim 10^{-4} \, h^3$ Mpc$^{-3}$, where $h = H_0/(100$
km s$^{-1}$ Mpc$^{-1})$ parameterizes Hubble's constant $H_0$), chosen
by considerations of observational efficiency.  For example, the
WiggleZ Survey obtained spectra of Emission Line Galaxies (Drinkwater
et al.\ 2010), whereas BOSS has instead focused on Luminous Red
Galaxies (Dawson et al.\ 2013).  Although these `single-tracer'
surveys have allowed increasingly precise tests of the cosmological
model -- recent examples of RSD growth-rate analyses include Blake et
al.\ (2011), Reid et al.\ (2012), Beutler et al.\ (2012), Samushia,
Percival \& Raccanelli (2012), Contreras et al.\ (2013) and de la
Torre et al.\ (2013) -- there are a number of potential advantages of
multiple-tracer surveys which we explore in this study.

First, surveying multiple populations of galaxies allows the key
assumptions needed to extract cosmological measurements to be examined
in an empirical way.  A fundamental systematic-error test is that our
cosmological conclusions should not depend on the galaxy population
used to trace the large-scale structure.  We flag in particular the
importance of modelling the {\it galaxy bias} which describes how
galaxy tracers populate the underlying large-scale structure.  When
using redshift-space distortions to measure the growth rate of
structure, $f = d \ln{\delta_m}/d \ln{a}$ in terms of the rate of
change in amplitude of a density perturbation $\delta_m$ with cosmic
scale factor $a$, it is common to assume that galaxy bias is linear
and deterministic, described by a single parameter $b$ which links the
galaxy and matter overdensities at position $\vec{x}$,
$\delta_g(\vec{x}) = b \, \delta_m(\vec{x})$.  In this case the
clustering anisotropy in redshift-space, i.e.\ the difference in the
amplitude of galaxy clustering as a function of the angle to the
line-of-sight, only depends on $f/b$.  However, in reality galaxy bias
is non-linear, scale-dependent and stochastic (e.g.\ Dekel \& Lahav
1999, Wild et al.\ 2005, Swanson et al.\ 2008, Cresswell \& Percival
2009, Marin 2011, Marin et al.\ 2013), and depends on the detailed
manner in which galaxies populate dark matter halos.  Comparison of
growth-rate measurements based on different galaxy tracers provides a
strict test of the modelling assumptions.

Secondly, a number of authors have pointed out that the availability
of multiple galaxy tracers across a volume of space allows improved
statistical errors in the measurements of certain cosmological
parameters (McDonald \& Seljak 2009, Seljak 2009, White, Song \&
Percival 2009, Gil-Marin et al.\ 2010, Bernstein \& Cai 2011, Hamaus,
Seljak \& Desjacques 2012, Abramo \& Leonard 2013).  These
improvements derive from the fact that, under the assumption of linear
galaxy bias, the tracers encode a common sample variance.  The
simplest example of this effect is to consider the overdensities in
two different galaxy populations which trace a single matter
overdensity: $\delta_{g,1}(\vec{x}) = b_1 \, \delta_m(\vec{x})$,
$\delta_{g,2}(\vec{x}) = b_2 \, \delta_m(\vec{x})$.  Neglecting all
other forms of noise, the ratio of these measured galaxy overdensities
allows the precise determination of $b_2/b_1$ independently of the
sample variance contained in $\delta_m$.

The next-simplest illustration, of particular relevance for our
analysis, is to consider measurements of the complex Fourier
amplitudes $\tilde{\delta}_{g,1}(\vec{k})$ and
$\tilde{\delta}_{g,2}(\vec{k})$ of the overdensity of two tracers for
the same wavevector $\vec{k}$, which has some angle to the
line-of-sight whose cosine is denoted by $\mu$.  In a linear model of
redshift-space distortions (Kaiser 1987):
\begin{align}
\tilde{\delta}_{g,1}(\vec{k}) &= (b_1 + f \mu^2) \, \tilde{\delta}_m(\vec{k}) \nonumber \\
\tilde{\delta}_{g,2}(\vec{k}) &= (b_2 + f \mu^2) \, \tilde{\delta}_m(\vec{k})
\label{eqmult}
\end{align}
where $\tilde{\delta}_m(\vec{k})$ is the corresponding (unknown)
Fourier amplitude of the underlying matter overdensity field, which
encodes the contribution of sample variance.  The ratio of these
measurements
\begin{equation}
\frac{\tilde{\delta}_{g,1}(\vec{k})}{\tilde{\delta}_{g,2}(\vec{k})} =
\frac{1 + \frac{f}{b_1} \, \mu^2}{\frac{b_2}{b_1} + \frac{f}{b_1} \,
  \mu^2}
\label{eqrat}
\end{equation}
in which we divide quantities on the right-hand side of the equation
by $b_1$ to clarify the observable combinations, does not contain the
unknown quantity $\tilde{\delta}_m(\vec{k})$, and is exactly known in
this idealized case.  By comparing measurements of this ratio at
different values of $\mu$, the quantities $b_2/b_1$, $f/b_1$ and
$f/b_2$ may be precisely determined.

There are a number of practical obstacles to realizing the advantages
outlined in the previous paragraph.  First, there is an additional
stochastic error component to equation (\ref{eqmult}), for example due
to galaxy `shot noise', that imposes a floor to the potential gains.
Hence multiple-tracer techniques demand high number-density galaxy
surveys in order to be effective.  Secondly, the expected gains scale
rapidly with the difference in galaxy bias factors, through the
strength of the variation of equation (\ref{eqrat}) with $\mu$ (with
no gain if $b_2 = b_1$).  The realization of this benefit conflicts
somewhat with the high number-density requirement, given that the
number density of dark matter halos rapidly diminishes with
increasing bias.  Also, although magnitude-limited galaxy surveys span
a wide range of galaxy luminosities (hence bias factors), there is
typically a strong luminosity-redshift correlation such that at a
given redshift the range of overlapping luminosities may be relatively
small.  Thirdly, equation (\ref{eqmult}) is only a good description of
galaxy clustering in the large-scale limit.  At smaller scales,
non-linear processes become increasingly important, weakening the
shared imprint of sample variance.  Fourthly, realistic survey
geometries render it impossible to measure directly the quantities of
equation (\ref{eqmult}): the underlying Fourier modes are convolved
with a survey selection function, such that measured power at some
wavevector depends on the underlying power at a range of different
wavevectors.

With all this said, the potential benefits of multiple-tracer surveys
are such that they are worth exploring in detail.  Indeed, although
there have been a number of studies of the theoretical implications of
the multiple-tracer technique, no analysis of data has yet been
presented.  In this study we remedy this gap by applying a
multiple-tracer power-spectrum analysis to one of the only high
number-density galaxy surveys at intermediate redshifts, the Galaxy
and Mass Assembly (GAMA) survey (Driver et al.\ 2011).  We explore the
resulting improvements in growth-rate measurements, and search for
systematic differences between results based on different galaxy
populations.

Our paper is structured as follows: section \ref{secsummary} provides
an overview of our implementation of the multiple-tracer method,
explaining how it differs from the illustrative equation (\ref{eqrat})
above.  Section \ref{secdata} describes the GAMA survey data, the
determination of the selection function, the clustering measurements
of different tracers and their covariances.  In section \ref{secmodel}
we fit redshift-space distortion models to these measurements and
compare the parameter fits resulting from single-tracer and
multiple-tracer analyses.  In section \ref{secsim} we validate our
investigations using mock catalogues derived from N-body simulations,
and in section \ref{secfish} we test our conclusions and compare with
other survey designs using Fisher matrix forecasts.  Section
\ref{secconc} summarizes our results.

\section{Overview of analysis method for correlated tracers}
\label{secsummary}

Before proceeding, we present an overview of our practical
implementation of the original insight of McDonald \& Seljak (2009).
First, rather than base our analysis on the 1-point statistics of the
density illustrated by equation (\ref{eqmult}), it is more convenient
to employ 2-point clustering statistics (we use the density power
spectrum).  In a Fisher-matrix sense, the 1-point statistics of
$(\tilde{\delta}_{g,1}, \tilde{\delta}_{g,2})$ and the 2-point
statistics described by the auto-power spectra and cross-power
spectrum of the two tracers $(P_1, P_2, P_c)$ contain identical
information.  Moreover, Fourier density modes may be binned when
measuring the power spectra, rendering the computation of a model
likelihood using the covariance matrix of the data tractable, given
the complicating effects of the realistic survey selection function.

Secondly, we avoid taking a ratio of observables such as equation
(\ref{eqrat}), even though this explicitly illustrates the removal of
sample variance.  Using a ratio in practice can lead to larger and
non-Gaussian errors, and the effects of the survey selection function
imply that the sample variance would not precisely cancel.  We instead
model the correlations between the tracer power spectra, induced by
the common sample variance, in the full covariance matrix of the
observables.

As a pedagogical illustration of our analysis method (see also
Bernstein \& Cai 2011) we consider auto-power spectrum measurements of
two tracers in a Fourier bin containing $M$ modes:
\begin{align}
P_1 &= (b_1 + f \mu^2)^2 \, P_m \, (1 + \alpha) + \epsilon_1 \nonumber \\
P_2 &= (b_2 + f \mu^2)^2 \, P_m \, (1 + \alpha) + \epsilon_2
\end{align}
where $P_m$ is the theoretical mean matter power spectrum in the bin,
which is assumed to be known exactly, $\alpha$ is the (single)
fluctuation from sample variance, which has a variance $\sigma^2 =
1/M$, and $(\epsilon_1, \epsilon_2)$ represent independent measurement
errors (e.g.\ from shot noise) such that $\langle \epsilon_1 \rangle =
\langle \epsilon_2 \rangle = \langle \epsilon_1 \epsilon_2 \rangle =
0$.  By analogy with equation (\ref{eqrat}) we consider estimating the
quantities $A = (b_1 + f \mu^2)^2$ and $B = (b_2 + f \mu^2)^2$.
Noting that $P_1/P_m$ is equal to the true value of $A$, plus the
independent fluctuations $A_{\rm true} \alpha + \epsilon_1/P_m$ (with
zero mean), the variance and covariance of the estimates of $A$ and
$B$ are
\begin{align}
\sigma_A^2 &= A^2 \sigma^2 + \langle \epsilon_1^2 \rangle/P_m^2 \nonumber \\
\sigma_B^2 &= B^2 \sigma^2 + \langle \epsilon_2^2 \rangle/P_m^2 \nonumber \\
\sigma_{AB}^2 &= A B \sigma^2
\end{align}
In the limit of small measurement error ($\langle \epsilon_i^2 \rangle
\rightarrow 0$) the fractional variances in $A$ and $B$ are both the
sample variance $\sigma^2$ -- but in this limit, the correlation
coefficient between $A$ and $B$, $\sigma_{AB}/\sqrt{\sigma_A
  \sigma_B}$, tends to unity.  The variance in the ratio $A/B$ is then
\begin{equation}
{\rm Var}(A/B) = (A/B)^2 \left[ \langle \epsilon_1^2 \rangle / A^2
  P_m^2 + \langle \epsilon_2^2 \rangle / B^2 P_m^2 \right]
\end{equation}
This contains no contribution from sample variance (is independent of
$\alpha$), reproducing the McDonald-Seljak result.

By using the power spectrum, rather than the density modes, we have
thrown away phase information.  But the reason that the
McDonald-Seljak method allows us to evade the sample variance limit is
that both tracers follow the same structure: thus a key aspect of the
method is that the phase of a given Fourier mode will be the same,
independent of tracer.  Since the power spectrum does not use this
fact, it may seem that we have not used the method properly and may
not suppress sample variance in the desired way.  In fact, the phase
adds no extra information to this particular analysis since it is part
of the sample variance that is cancelled in any case when forming the
original ratio in equation (\ref{eqrat}).

In the equations above we have just considered the two auto-power
spectra, $P_1 = |\tilde{\delta}_1|^2$ and $P_2 =
|\tilde{\delta}_2|^2$.  What is the role of the cross-power spectrum
$P_c = {\rm Re}\{ \tilde{\delta}_1 \, \tilde{\delta}_2^* \}$?  In our
above model, this would be:
\begin{equation}
P_c = (b_1 + f \mu^2) \, (b_2 + f \mu^2) \, P_m \, (1 + \alpha) +
\epsilon_c
\end{equation}
When the measurement errors are small, $P_c = \sqrt{P_1 \, P_2}$ and
there is no extra information in the cross-power spectrum.  However,
determination of the cross-power spectrum does provide some
independent validation of the underlying assumption of close
correlation between the two tracers (e.g., scrambling the phases of
$\tilde{\delta}_1$ and $\tilde{\delta}_2$ would leave the auto-power
spectrum measurements unchanged, but yield zero cross-power) and,
furthermore, serves to test the assumption of linear galaxy bias.  We
note that for datasets where the measurement errors $\epsilon$ are not
negligible, the cross-power spectrum adds information to the parameter
determinations.  In section \ref{secfish} we use a full Fisher matrix
analysis to consider this point further.

The measurement errors for the GAMA dataset analyzed in this study are
sufficiently small that the cross-power spectrum adds negligible
information (improving the determination of the growth rate by only
$0.2\%$ according to the Fisher matrix forecasts presented in section
\ref{secfish} below).  Indeed, its inclusion in the primary analysis
causes technical difficulties with inverting the relevant covariance
matrices, which are nearly singular.  Therefore, although for
completeness we present full derivations of the covariances including
the cross-power spectrum, we restrict our parameter fits to the
auto-power spectra of the two tracers, and use the cross-power
spectrum solely for validation of the method.

\section{Power spectrum data}
\label{secdata}

\subsection{GAMA survey}

The Galaxy and Mass Assembly (GAMA) project (Driver et al.\ 2011) is a
multi-wavelength photometric and spectroscopic survey.  The redshift
survey, which has been carried out with the Anglo-Australian Telescope
(AAT), has provided a dense, highly-complete sampling of large-scale
structure up to redshift $z \sim 0.5$.  The primary target selection
is $r < 19.8$ (where $r$ is an extinction-corrected SDSS Petrosian
magnitude).

In this study we analyzed a highly-complete subsample of the latest
survey dataset, known as the GAMA II equatorial fields.  This
subsample covers three $12 \times 5$ deg regions centred at 09h, 12h
and 14h30m which we refer to as G09, G12 and G15, respectively.  The
GAMA I target selection is described by Baldry et al.\ (2010) and GAMA
II in Liske et al.\ (in preparation).  For GAMA II, the fields were
widened by 1 degree and the $r$-band selection magnitude was changed
from SDSS DR6 to DR7 (updated to ubercalibration, Abazajian et
al.\ 2009).  We restricted the input catalogue to $r < 19.8$ and only
included targets that satisfied the $r$-band star-galaxy separation;
this excluded some $J-K$ selection because the near-IR photometry had
significant missing coverage.  We obtained the GAMA~II data from {\tt
  TilingCatv41}, selecting 185052 targets ({\tt SURVEY\_CLASS} $\ge
5$).

\begin{figure}
\begin{center}
\resizebox{8cm}{!}{\rotatebox{270}{\includegraphics{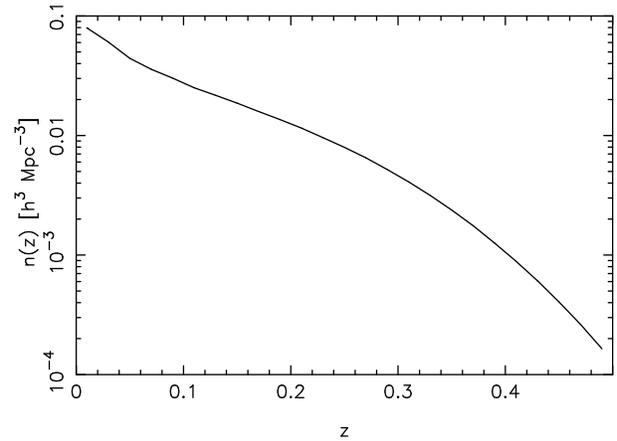}}}
\end{center}
\caption{The average number density of GAMA galaxies as a function of
  redshift.  This plot is constructed by combining data in the three
  survey regions.}
\label{figndens}
\end{figure}

Papers based on GAMA I data had used redshifts obtained from a
semi-automatic code, {\tt runz}, involving some user interaction.  The
redshifts for GAMA II ({\tt TilingCatv41}), used here, have been
updated using a fully automatic cross-correlation code that can
robustly measure absorption and emission line redshifts (Baldry et
al.\ in preparation). This significantly improved the reliability of
the measured redshifts from the AAT.  We restricted the redshift
catalogue to galaxies with `good' redshifts ({\tt NQ} $\ge 3$) in the
range $0.002 < z < 0.5$.  In the (G09, G12, G15) regions we utilized
$(57194, 61278, 60107)$ galaxies in our analysis.  K-corrections were
calculated with {\tt kcorrect\_v4.2} (Blanton \& Roweis 2007) using
SDSS model magnitudes (see Loveday et al.\ 2012 for more details).
Fig.\ \ref{figndens} displays the average number density of these GAMA
galaxies as a function of redshift, illustrating the high values
available for our analysis, which exceed $10^{-2} \, h^3$ Mpc$^{-3}$
in the range $z < 0.25$.

\begin{figure}
\begin{center}
\resizebox{8cm}{!}{\rotatebox{270}{\includegraphics{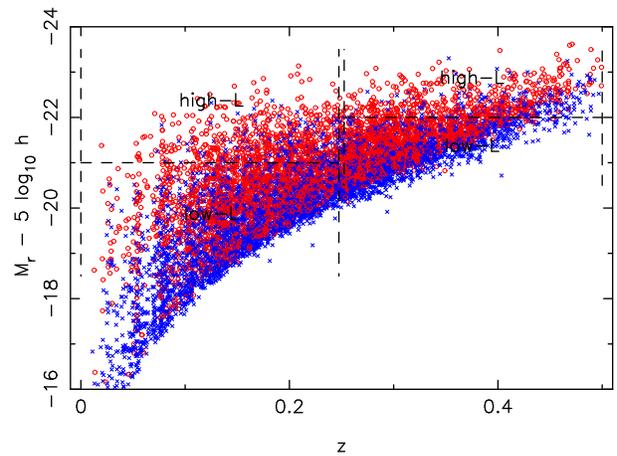}}}
\end{center}
\caption{The distribution of absolute magnitudes $M_r$ and redshifts
  $z$ for the GAMA galaxies used in our analysis, which satisfy the
  selection criteria described in the text.  The blue and red colour
  subsamples are plotted as crosses and open circles, respectively,
  and illustrated by appropriate colouring of the data points.  The
  `high-$L$' and `low-$L$' subsamples are shown by the ranges
  indicated on the figure.  In this plot, the galaxies have been
  randomly subsampled by a factor of 20, for clarity.}
\label{figlumz}
\end{figure}

We performed clustering measurements of galaxies in two independent
redshift ranges $0 < z < 0.25$ and $0.25 < z < 0.5$.  For each
redshift range we split the data into two subsamples in order to apply
multiple-tracer techniques.  We considered splits by colour and
luminosity.  First, we divided galaxies into two colour classes, `red'
and `blue', using a redshift-dependent division in the observed colour
\begin{equation}
g - i = 0.8 + 3.2 \, z
\end{equation}
which traces a clear bimodality in the observed GAMA colour
distribution at all redshifts.  Here, $g$ and $i$ are model magnitudes
in the appropriate bands.  Alternatively, we explored splitting
galaxies into two luminosity classes based on the rest-frame absolute
magnitude in the $r$-band.  For the redshift ranges $(0 < z < 0.25,
0.25 < z < 0.5)$ we take these luminosity divisions at $M_r - 5 \,
{\rm log}_{10} h = (-21, -22)$.  Fig.\ \ref{figlumz} illustrates the
luminosity-redshift distribution of GAMA galaxies, colour-coded to
indicate galaxies selected as `red' and `blue'.

\subsection{Survey selection function}
\label{secsel}

In order to quantify the GAMA galaxy clustering, we must first define
the survey selection function which describes the expected galaxy
distribution in the absence of clustering.  We separated this
selection function into independent angular and radial components.

\begin{figure}
\begin{center}
\resizebox{8cm}{!}{\rotatebox{0}{\includegraphics{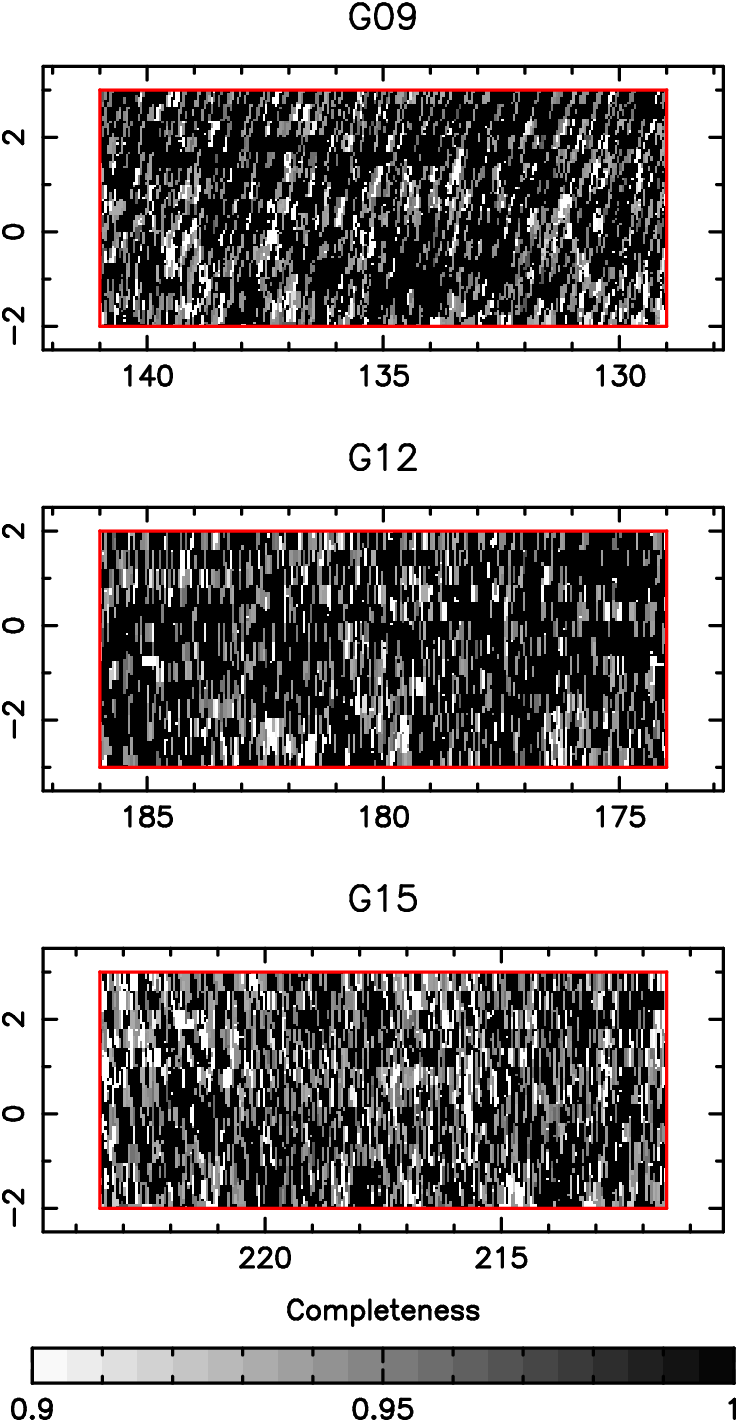}}}
\end{center}
\caption{The angular completeness maps for each of the GAMA regions
  analyzed in this study.  The $x$- and $y$-axes correspond to
  R.A. and Dec. co-ordinates, respectively, in degrees.}
\label{figangcomp}
\end{figure}

The angular selection function for each GAMA region describes the
exact sky coverage of the input target imaging catalogues, together
with the small fluctuations in the redshift completeness of the
spectroscopic follow-up.  We used the masks and software available in
the survey database, {\tt
  completeness\_maps:software:mask\_redshift\_r} and {\tt
  completeness\_maps:software:mask\_sdss}, to produce angular
completeness maps in (R.A., Dec.) on a fine pixel grid.  These maps
are displayed in Fig.\ \ref{figangcomp}, in which we note the very
high level of redshift completeness across each survey region, with a
mean value of $97\%$.

\begin{figure}
\begin{center}
\resizebox{8cm}{!}{\rotatebox{270}{\includegraphics{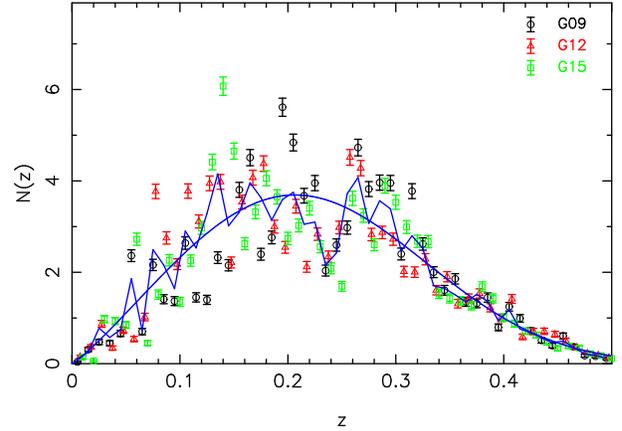}}}
\end{center}
\caption{Determination of the radial selection function for all GAMA
  galaxies in our sample in the redshift interval $0 < z < 0.5$.  The
  figure shows the redshift distributions within each of the three
  GAMA regions (black circles, red triangles and green squares)
  together with the combined $N(z)$ (jagged blue solid line) and
  fitted model (smooth blue solid line).  The $y$-axis is normalized
  such that $\int N(z) \, dz = 1$.  The plotted error bars are double
  the Poisson error predicted from the number of counts in each bin,
  noting that this is sub-dominant to the region-to-region
  fluctuations.  At higher redshifts $z > 0.35$ the extra available
  cosmic volume results in these fluctuations becoming less
  significant.}
\label{fignz}
\end{figure}

We determined the radial selection function of a given colour or
luminosity subsample using an empirical smooth fit to the observed
galaxy redshift distribution $N(z)$ of that subsample.  Measurements
of $N(z)$ in individual GAMA regions contain significant fluctuations;
we reduced this by combining the 3 regions.  We found that the model
\begin{equation}
N(z) \propto \left( \frac{z}{z_0} \right)^\alpha \, e^{-(z/z_0)^\beta}
\end{equation}
provided a good fit to all the relevant redshift distributions in
terms of the 3 parameters $(z_0, \alpha, \beta)$.  Fig.\ \ref{fignz}
displays an example of this model fitted to all GAMA galaxies in our
sample (normalized such that $\int N(z) \, dz = 1$).

\begin{figure*}
\begin{center}
\resizebox{16cm}{!}{\rotatebox{270}{\includegraphics{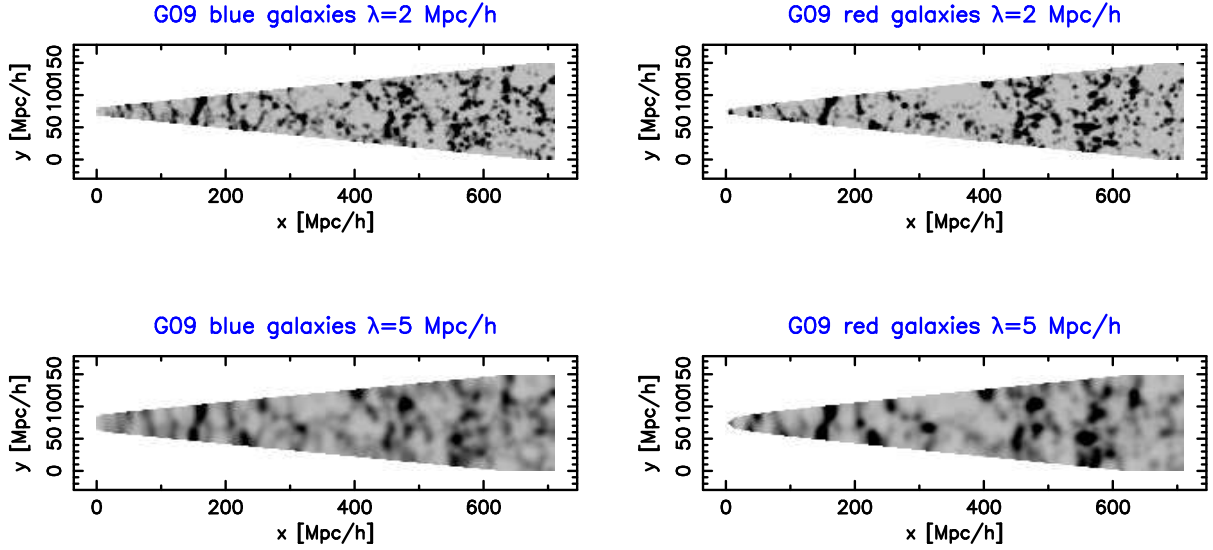}}}
\end{center}
\caption{The galaxy overdensity field within the G09 region,
  determined from the gridded data and selection function, and
  projected onto a 2D plane parallel to the line-of-sight (such that
  the $x$-, $y$- and $z$-axes are oriented in the redshift, right
  ascension and declination directions, respectively).  The left-hand
  and right-hand panels show the measurements for blue and red
  galaxies, respectively.  The top and bottom rows illustrate two
  choices of smoothing scale, $2$ and $5 \, h^{-1}$ Mpc.
  Qualitatively, it can be seen that the different galaxy subsamples
  are tracing the same underlying large-scale structure.}
\label{figdens}
\end{figure*}

Using the survey selection function we can visualize the galaxy
overdensity field within each region.  For the purposes of this
calculation we binned the galaxy distribution and normalized selection
function in a 3D co-moving co-ordinate grid, denoting these gridded
distributions as $D$ and $R$, and then determined the overdensity
field $\delta$ by smoothing these distributions with a Gaussian kernel
$G(\vec{x}) = e^{-(\vec{x}.\vec{x})/2\lambda^2}$ such that $\delta =
{\rm smooth}(D)/{\rm smooth}(R) - 1$ and $\langle \delta \rangle = 0$.

\begin{figure}
\begin{center}
\resizebox{8cm}{!}{\rotatebox{270}{\includegraphics{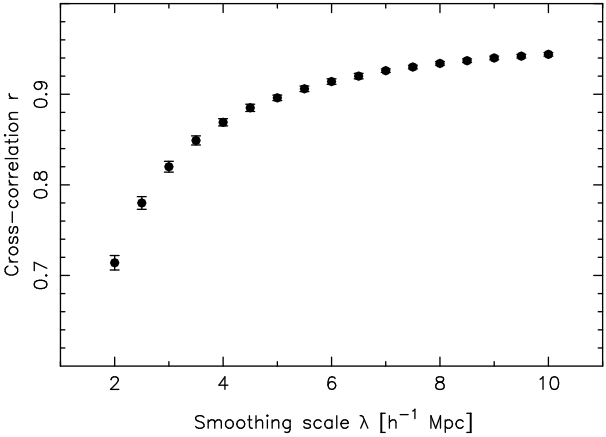}}}
\end{center}
\caption{The cross-correlation coefficient in configuration space
  between the red and blue galaxy overdensity fields, as a function of
  the smoothing length $\lambda$ of a Gaussian kernel.  These
  measurements correspond to a redshift interval $0 < z < 0.25$
  combining all 3 GAMA regions, and illustrate the high level of
  correlated sampled variance in the GAMA galaxy subsamples.}
\label{figcrosscorr}
\end{figure}

Using the $0 < z < 0.25$ redshift interval of the G09 region for
illustration, Fig.\ \ref{figdens} compares the smoothed galaxy density
fields determined from the blue and red galaxy subsamples for $\lambda
= 2$ and $5 \, h^{-1}$ Mpc, illustrating that qualitatively these two
populations are tracing the same underlying large-scale structure.
Fig.\ \ref{figcrosscorr} quantifies this observation by measuring the
cross-correlation coefficient between the red and blue galaxy
overdensity fields $r = \langle \delta_1 \delta_2 \rangle /
\sqrt{\langle \delta_1^2 \rangle \langle \delta_2^2 \rangle}$ as a
function of the smoothing scale $\lambda$.  We again analyzed a $0 < z
< 0.25$ redshift slice, computing the cross-correlation coefficient
over all 3 survey regions.  The errors in the measurements were
determined by jack-knife methods (using 100 jack-knife partitions per
survey region).  The cross-correlation coefficient rises to $r > 0.9$
on scales $\lambda > 5 \, h^{-1}$ Mpc, dropping on smaller scales due
to the effects of shot noise and the manner in which different classes
of galaxy populate dark matter halos (scale-dependent and/or
stochastic galaxy bias).  These analyses illustrate the strong level
of correlated sample variance in the multiple GAMA galaxy populations;
in the next section we quantify these effects using power-spectrum
measurements.

\subsection{Power spectrum measurements}

We measured the power spectra of GAMA galaxies within each separate
survey region, fitted models to these measurements, and combined the
results of the fits assuming each region was independent.  Our
measurements of the auto-power and cross-power spectra of galaxies
within each GAMA region were based on the optimal-weighting estimation
scheme of Feldman, Kaiser \& Peacock (FKP; 1994), which we generalized
to cross-power spectra (also see Smith 2009).

First we converted the galaxy distribution in a particular region to
co-moving co-ordinates, assuming a fiducial flat $\Lambda$CDM
cosmology with matter density $\Omega_{\rm m} = 0.27$.  We then
enclosed the survey cone within the relevant redshift interval by a
cuboid of sides $(L_x, L_y, L_z)$ with volume $V = L_x L_y L_z$, and
gridded the galaxy catalogue in cells numbering $(n_x, n_y, n_z)$
using nearest grid point assignment to produce distributions
$N_1(\vec{x})$ and $N_2(\vec{x})$ for the two tracers.  The cell
dimensions were chosen such that the Nyquist frequencies in each
direction (e.g.\ $k_{{\rm Nyq},x} = \pi n_x/L_x$) exceeded the maximum
frequency of measured power by a factor of at least four.

We then applied a Fast Fourier transform to the gridded data,
weighting each pixel by factors $w_1(\vec{x})$ and $w_2(\vec{x})$ for
the two tracers, respectively:
\begin{equation}
{\rm FFT}(N_{w,\alpha}) \equiv \tilde{N}_{w,\alpha}(\vec{k}) = \sum_{\vec{x}} w_\alpha(\vec{x}) \, N_\alpha(\vec{x}) \, e^{i\vec{k}.\vec{x}}
\end{equation}
where $\alpha = 1$ or $2$ labels the galaxy population in all
equations in this section, and the weighting factors are given by
\begin{equation}
w_\alpha(\vec{x}) = \frac{1}{1 + W_\alpha(\vec{x}) \, N_c \, n_\alpha \, P_0}
\label{eqweight}
\end{equation}
In equation (\ref{eqweight}), $N_c = n_x n_y n_z$ is the total number
of grid cells, $n_\alpha$ is the mean number density of each set of
tracers, and $P_0 = 5000 \, h^{-3}$ Mpc$^3$ is a characteristic value
of the power spectrum at the scales of interest ($k \sim 0.1 \, h$
Mpc$^{-1}$; we note that this can be generalized as a function of
luminosity following Percival, Verde \& Peacock (2004), which is
beyond the scope of the current study).  $W_\alpha(\vec{x})$ is
proportional to the survey selection function at each grid cell
determined in section \ref{secsel}, normalized such that
$\sum_{\vec{x}} W_\alpha(\vec{x}) = 1$.

We note that the application of FKP weighting to multiple-tracer
analyses requires caution: this weighting is designed to minimize the
error in the measured power spectrum by balancing the effects of
sample variance and shot noise, and yet (in the ideal case) the sample
variance error is suppressed by the combination of the two tracers.
However, for realistic surveys with a selection function and shot
noise, the sample variance is only partially suppressed.  We repeated
our analyses for different choices of $P_0$: for no weighting ($P_0 =
0$) we found that the error in the measured growth rate in the various
cases increased by 30-40$\%$, whereas doubling the characteristic
power to $P_0 = 10{,}000 \, h^{-3}$ Mpc$^3$ produced a result almost
identical to the fiducial choice of $P_0 = 5000 \, h^{-3}$ Mpc$^3$.
We may also be concerned that the slightly different weights ($w_1 \ne
w_2$) applied to each subsample, owing to their different selection
functions in equation (\ref{eqweight}), may undermine the correlated
sample variance and weaken the eventual growth rate determination.  In
order to test this concern we repeated the power spectrum measurements
applying an identical weight to each subsample equal to $(w_1+w_2)/2$.
We found that the error in the final growth rate was unchanged
compared to our default implementation.  Finally, we note that more
sophisticated mass-dependent weighting schemes have been proposed by
some authors (Seljak, Hamaus \& Desjacques 2009; Cai, Bernstein \&
Sheth 2011); these will be considered in future work.

\begin{figure*}
\begin{center}
\resizebox{13cm}{!}{\rotatebox{270}{\includegraphics{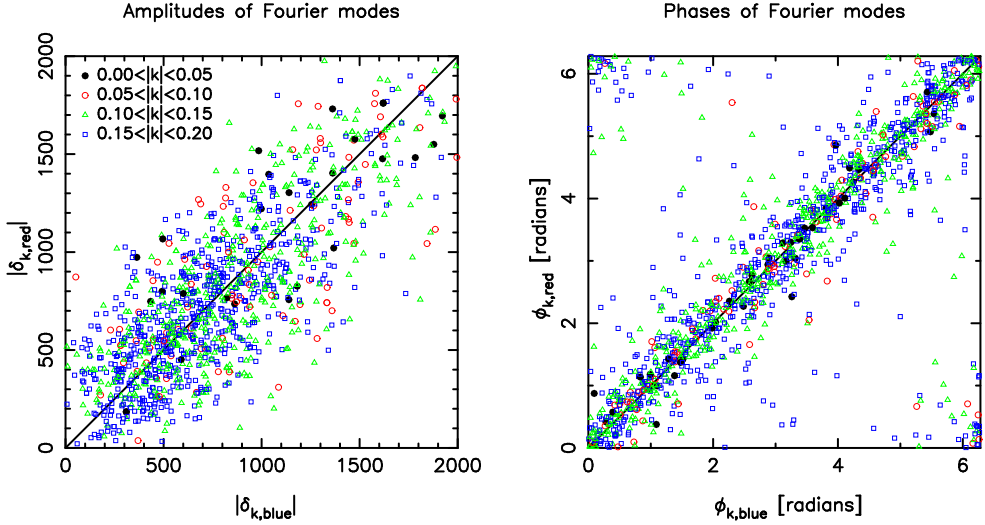}}}
\end{center}
\caption{A mode-by-mode comparison of the moduli
  $|\tilde{\delta}(\vec{k})|$ and phases $\phi(\vec{k})$ of the
  complex Fourier amplitudes $\tilde{\delta}(\vec{k})$ estimated for
  the red and blue galaxy subsamples for the redshift interval $0 < z
  < 0.25$.  The points are coded by plotting symbol and colour into
  four wavenumber bins in the range $k < 0.2 \, h$ Mpc$^{-1}$.  We
  note the strong correlations between the measurements of a given
  Fourier amplitude for the two different tracers.  In the right-hand
  panel, the data points which appear in the upper-left and
  lower-right corners result from the $2\pi$ wrapping of the phases
  and support the correlation.}
\label{figdkcomp}
\end{figure*}

We measured the complex Fourier amplitudes of the two tracers as
\begin{equation}
\tilde{\delta}_\alpha(\vec{k}) = \tilde{N}_{w,\alpha}(\vec{k}) -
N_\alpha \, \tilde{W}_{w,\alpha}(\vec{k})
\end{equation}
where $N_\alpha$ is the total number of galaxies for population
$\alpha$, and $\tilde{W}_{w,\alpha}$ is the Fast Fourier transform of
the weighted selection function
\begin{equation}
{\rm FFT}(W_{w,\alpha}) \equiv \tilde{W}_{w,\alpha}(\vec{k}) =
\sum_{\vec{x}} w_\alpha(\vec{x}) \, W_\alpha(\vec{x}) \,
e^{i\vec{k}.\vec{x}}
\end{equation}
Fig.\ \ref{figdkcomp} compares the moduli $|\tilde{\delta}_\alpha|$
and phases $\phi_\alpha$ of the complex Fourier amplitudes
$\tilde{\delta}_\alpha = |\tilde{\delta}_\alpha| \, e^{i \phi_\alpha}$
for the red and blue galaxy subsamples for the $0 < z < 0.25$ redshift
interval.  The common sample variance induces clear correlations
between the moduli and phases of the different populations.

\begin{figure*}
\begin{center}
\resizebox{13cm}{!}{\rotatebox{270}{\includegraphics{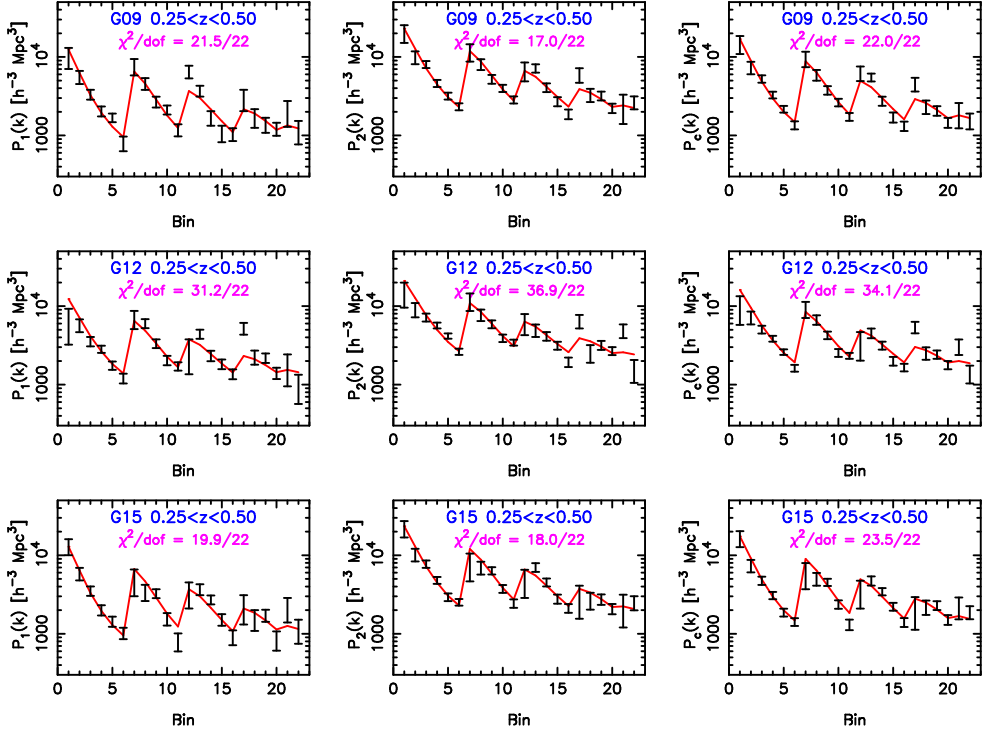}}}
\end{center}
\caption{Measurements of the two auto-power spectra and cross-power
  spectrum for the blue and red subsamples of GAMA galaxies in the
  redshift interval $0.25 < z < 0.5$.  The columns (from
  left-to-right) correspond to $(P_1, P_2, P_c)$.  The rows (from
  top-to-bottom) display the measurements for the 3 regions (G09, G12,
  G15).  The solid line is the best-fitting model (to the 2 auto-power
  spectra).  The data points are ordered by looping over the bins of
  $k_\perp$ and then $k_\parallel$, only plotting bins for which $|k|
  = \sqrt{k_\perp^2 + k_\parallel^2} < 0.3 \, h$ Mpc$^{-1}$,
  constituting 22 bins.  The $x$-axis represents the bin number in the
  ordering, and the `saw-tooth' pattern is produced by the repeated
  looping over $k_\perp$.  The values of the $\chi^2$ statistic for
  the model are quoted separately for each power spectrum and region.}
\label{figpkmeas}
\end{figure*}

In the Appendix we derive the estimators of the two auto-power
spectra, $P_1(\vec{k})$ and $P_2(\vec{k})$, and cross-power spectrum
$P_c(\vec{k})$.  The final expressions are:
\begin{align}
\hat{P}_\alpha(\vec{k}) &= \frac{V \, \left[ |\tilde{\delta}_\alpha(\vec{k})|^2 - N_\alpha \sum_{\vec{x}} W_\alpha \, w_\alpha^2 \right]}{N_c \, N_\alpha^2 \sum_{\vec{x}} W_\alpha^2 \, w_\alpha^2} \nonumber \\
\hat{P}_c(\vec{k}) &= \frac{V \, {\rm Re} \left\{ \tilde{\delta}_1(\vec{k}) \, \tilde{\delta}_2^*(\vec{k}) \right\} }{N_c \, N_1 \, N_2 \sum_{\vec{x}} W_1 \, w_1 \, W_2 \, w_2 }
\label{eqpkest}
\end{align}
We note that the expectation values of the estimators in equation
(\ref{eqpkest}) are a convolution of the underlying model power
spectra:
\begin{align}
\langle \hat{P}_\alpha(\vec{k}) \rangle &= \frac{V^3}{(2\pi)^3} \int P_\alpha(\vec{k}') \, |\tilde{n}_{w,\alpha}(\delta \vec{k})|^2 \, d^3\vec{k}' \nonumber \\
\langle \hat{P}_c(\vec{k}) \rangle &= \frac{V^3}{(2\pi)^3} \int P_c(\vec{k}') \, {\rm Re} \left\{ \tilde{n}_{w,1}(\delta \vec{k}) \, \tilde{n}_{w,2}^*(\delta \vec{k}) \right\} \, d^3\vec{k}'
\label{eqpkconv}
\end{align}
where $\tilde{n}_{w,\alpha} = N_\alpha \, \tilde{W}_{w,\alpha}$ and
$\delta \vec{k} = \vec{k}'-\vec{k}$.  We averaged the power spectrum
amplitudes for the different Fourier modes in bins of wavevector
perpendicular and parallel to the line-of-sight, $(k_\perp,
k_\parallel)$.  Since in our analysis we orient the $x$-axis parallel
to the line-of-sight to the centre of each survey region, and each
region has a narrow and deep geometry, we can make the flat-sky
approximation $k_\perp = \sqrt{k_y^2 + k_z^2}$, $k_\parallel = |k_x|$
(noting that any resulting systematic distortion is negligible
compared with the sample-variance error in our measurements).  We used
wavevector bins of width $\Delta k_\perp = \Delta k_\parallel = 0.05
\, h$ Mpc$^{-1}$ in the analysis, only considering bins for which $|k|
= \sqrt{k_\perp^2 + k_\parallel^2} < 0.3 \, h$ Mpc$^{-1}$ because of
concerns over modelling non-linearities in the power spectrum at
smaller scales, which are explored further in section \ref{secmodel}.
We also excluded the largest-scale (lowest) bin in $k_\parallel$, $0 <
k_\parallel < 0.05 \, h$ Mpc$^{-1}$, whose measured power is prone to
systematic effects from the radial selection function fits.  The final
result was a total of 22 bins.  Fig.\ \ref{figpkmeas} displays the
binned auto-power and cross-power spectrum measurements for the blue
and red galaxy subsamples in the redshift interval $0.25 < z < 0.5$,
for each of the three GAMA regions.

\begin{figure*}
\begin{center}
\resizebox{13cm}{!}{\rotatebox{270}{\includegraphics{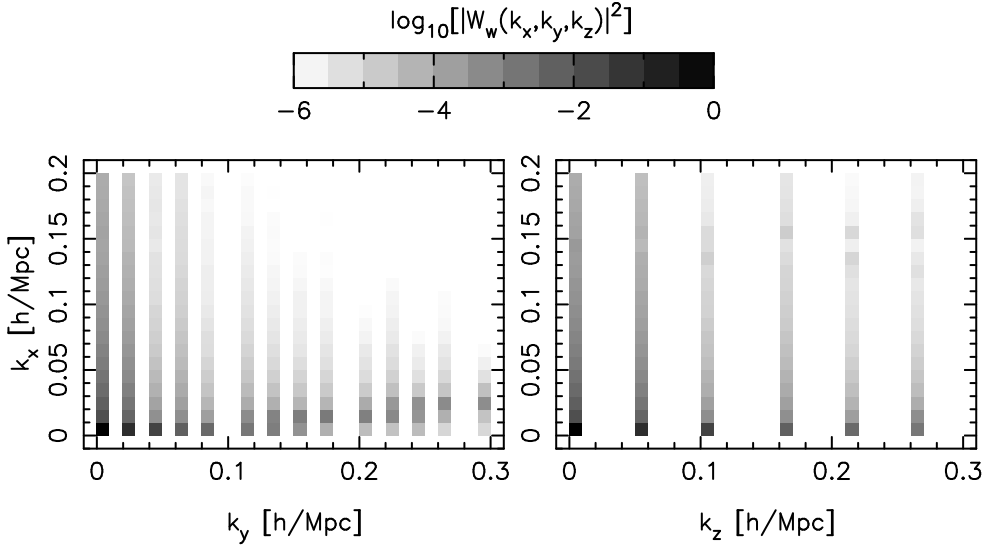}}}
\end{center}
\caption{The structure of the Fourier transform of the weighted survey
  selection function, $|\tilde{W}_w(\vec{k})|^2$, which determines the
  relative weighting of the power spectrum modes combined by the
  convolution of equation (\ref{eqpkconv}).  As an example we display
  the selection function of the `blue' subsample for the $0.25 < z <
  0.5$ redshift slice of the G09 region.  We show 2D projections of
  this function in the space of $(k_y,k_x)$ and $(k_z,k_x)$, where the
  $x$-axis is oriented along the line-of-sight and the $y$- and
  $z$-axes are parallel to the (long) right ascension and (short)
  declination directions, respectively.  The restricted $y$- and
  $z$-dimensions of the survey cuboid imprint a series of diminishing
  peaks with a regular spacing $\Delta k = 2\pi/L$.  The amplitude of
  the function is indicated by the greyscale shown in the upper
  legend, note the logarithmic scale.}
\label{figwk}
\end{figure*}

Fig.\ \ref{figwk} displays an example of the structure of the Fourier
transform of the weighted selection function,
$|\tilde{W}_w(\vec{k})|^2$, which determines the relative weighting of
the power spectrum modes combined by the convolution of equation
(\ref{eqpkconv}).  As expected, this function contains a series of
diminishing peaks along each axis spaced by $\Delta k = 2\pi/L$, in
accordance with the dimension $L$ of the survey cuboid parallel to
that axis.  These peaks are hence particularly widely-spaced parallel
to the narrow, declination direction of the survey geometry.  This
structure was fully modelled in our parameter fits.  When fitting
models, we re-cast the convolution integrals of equation
(\ref{eqpkconv}) as matrix multiplications for reasons of numerical
speed:
\begin{align}
\langle \hat{P}_\alpha(i) \rangle &= \sum_j (M_\alpha)_{ij} \, P_{{\rm mod},\alpha}(j) \nonumber \\
\langle \hat{P}_c(i) \rangle &= \sum_j (M_c)_{ij} \, P_{{\rm mod},c}(j)
\end{align}
where $(P_{{\rm mod},1}, P_{{\rm mod},2}, P_{{\rm mod},c})$ are the
model auto-power and cross-power spectra for the 2 populations,
evaluated at the centres of the Fourier bins.  We determined the
convolution matrices $(M_1, M_2, M_c)$ by evaluating the full
integrals given in equation (\ref{eqpkconv}) for a set of unit model
vectors, and tested that this produced a negligible change in results
compared to implementing the full convolution.

We defined the effective redshift of each power spectrum measurement
by weighting each pixel in the 3D selection function by its
contribution to the power spectrum error:
\begin{equation}
z_{\rm eff}(k) = \sum_{\vec{x}} z \times \left[ \frac{n(\vec{x}) \,
    P(k)}{1 + n(\vec{x}) \, P(k)} \right]^2
\end{equation}
We evaluated this relation at $k = 0.1 \, h$ Mpc$^{-1}$ (although the
results do not depend strongly on this choice).  The effective
redshifts of the measurements in the redshift intervals $(0 < z <
0.25, 0.25 < z < 0.5$ are $z_{\rm eff} = (0.18, 0.38)$, with a very
weak dependence on galaxy type.

\subsection{Covariance matrix}

The survey selection functions and correlated sample variance induce
covariances between the estimates of the two auto-power and
cross-power spectra of the galaxy populations for two Fourier modes
$\vec{k}$ and $\vec{k}'$.  These covariances are derived in the
Appendix; the expressions for the auto-power spectrum follow Feldman
et al.\ (1994), to our knowledge the other formulae are new (regarding
the inclusion of the selection function and weights, but also see
Smith 2009).  The results may be conveniently expressed in terms of
the functions
\begin{align}
Q_\alpha(\vec{x}) &= w_\alpha^2(\vec{x}) \, n_\alpha^2(\vec{x}) \nonumber \\
Q_c(\vec{x}) &= w_1(\vec{x}) \, n_1(\vec{x}) \, w_2(\vec{x}) \, n_2(\vec{x}) \nonumber \\
S_\alpha(\vec{x}) &= w_\alpha^2(\vec{x}) \, n_\alpha(\vec{x})
\end{align}
Further defining
\begin{align}
z_\alpha(\vec{k},\vec{k}') &= P_\alpha(\vec{k}) \, \tilde{Q}_\alpha(\vec{k}'-\vec{k}) + \tilde{S}_\alpha(\vec{k}'-\vec{k}) \nonumber \\
z_c(\vec{k},\vec{k}') &= P_c(\vec{k}) \, \tilde{Q}_c(\vec{k}'-\vec{k})
\end{align}
the equations for the covariances in this approximation are
\begin{align}
\langle \delta \hat{P}_\alpha(\vec{k}) \, \delta \hat{P}_\alpha(\vec{k}') \rangle &= \frac{|z_\alpha(\vec{k},\vec{k}')|^2}{\tilde{Q}_\alpha(0)^2} \nonumber \\
\langle \delta \hat{P}_1(\vec{k}) \, \delta \hat{P}_2(\vec{k}') \rangle &= \frac{|z_c(\vec{k},\vec{k}')|^2}{\tilde{Q}_1(0) \, \tilde{Q}_2(0)} \nonumber \\
\langle \delta \hat{P}_\alpha(\vec{k}) \, \delta \hat{P}_c(\vec{k}') \rangle &= \frac{ {\rm Re} \left\{ z_\alpha(\vec{k},\vec{k}') \, z_c(\vec{k},\vec{k}')^* \right\} }{\tilde{Q}_\alpha(0) \, \tilde{Q}_c(0)} \nonumber \\
\langle \delta \hat{P}_c(\vec{k}) \, \delta \hat{P}_c(\vec{k}') \rangle &= \frac{ |z_c(\vec{k},\vec{k}')|^2 + {\rm Re} \left\{ z_1(\vec{k},\vec{k}') \, z_2(\vec{k},\vec{k}')^* \right\} }{2 \, \tilde{Q}_c(0)^2}
\label{eqcov}
\end{align}
where $\delta \hat{P} = \hat{P} - \langle \hat{P} \rangle$.  The
derivation of these covariance relations involves the following
approximations (Feldman et al.\ 1994):
\begin{itemize}
\item The Fourier coefficients $\tilde{\delta}(\vec{k})$ are
  Gaussian-distributed, such that the 4-point function assumes a
  simple form (derived in the Appendix).
\item The galaxy distribution forms a Poisson sample of the density
  field.
\item The power spectrum is effectively constant over the coherence
  scale defined by the Fourier transform of the survey selection
  function.
\end{itemize}

It is a useful cross-check of these equations to consider the special
case of a uniform selection function and weights.  In this case
$Q_\alpha = n_\alpha^2$, $Q_c = n_1 n_2$ and $S_\alpha = n_\alpha$,
and the equations simplify to:
\begin{align}
\langle \delta \hat{P}_\alpha \, \delta \hat{P}_\alpha \rangle &= \left( P_\alpha + \frac{1}{n_\alpha} \right)^2 \nonumber \\
\langle \delta \hat{P}_1 \, \delta \hat{P}_2 \rangle &= P_c^2 \nonumber \\
\langle \delta \hat{P}_\alpha \, \delta \hat{P}_c \rangle &= P_c \left( P_\alpha + \frac{1}{n_\alpha} \right) \nonumber \\
\langle \delta \hat{P}_c \, \delta \hat{P}_c \rangle &= \frac{1}{2} \left[ P_c^2 + \left( P_1 + \frac{1}{n_1} \right) \left( P_2 + \frac{1}{n_2} \right) \right]
\label{eqcovsimp}
\end{align}
In physical terms, the covariance within each auto-power spectrum is
driven by a combination of sample variance ($P_\alpha$) and shot noise
($1/n_\alpha$).  The covariance between different auto-power spectra
does not involve shot noise (since each galaxy can only appear in one
subsample) but depends on sample variance via the cross-power spectrum
($P_c$).  Covariances involving the cross-power spectrum are more
complicated, involving both sample variance and shot noise
contributions.

We note an important technical difficulty that arises when inverting
the full covariance matrices for the two auto power-spectra and cross
power-spectrum.  In the approximation of linear bias and common
non-linear RSD damping (see section \ref{secmodel}), the cross-power
spectrum is a simple geometric mean of the two auto-power spectra:
$P_c = \sqrt{P_1 \, P_2}$.  In the limit of high galaxy number
density, such that shot noise is negligible, the cross-power spectrum
measurement then adds no information to that already present in the
two auto-power spectra.  We can verify this mathematically by taking
the limit of equation (\ref{eqcovsimp}) as $n_\alpha \rightarrow
\infty$.  The covariance matrix for the measurement of $(P_1, P_2,
P_c)$ for a single Fourier mode becomes:
\begin{equation}
C(\vec{k}) = \left( \begin{array}{ccc}
P_1^2 & P_c^2 & P_1 \, P_c \\
P_c^2 & P_2^2 & P_2 \, P_c \\
P_1 \, P_c & P_2 \, P_c & \frac{1}{2} \left( P_c^2 + P_1 \, P_2 \right)
\end{array} \right)
\end{equation}
which, given that $P_c^2 = P_1 \, P_2$, implies that $|C| = 0$ and the
matrix is singular.  (The fact that the cross-power spectrum adds no
information as $n_\alpha \rightarrow \infty$ is also demonstrated
later by the Fisher matrix calculations in section \ref{secfish}).

The number densities of the GAMA multiple-tracer populations are well
within the regime where the contribution of the cross-power spectrum
to the parameter constraints is negligible, and in fact we found that
the full covariance matrix was not always positive definite (even for
finite $n_1$ and $n_2$).  We traced the cause of this issue as the
approximation made in equations (\ref{eqapprox1}) and
(\ref{eqapprox2}) which results in the covariance matrix of equation
(\ref{eqcov}); evaluating instead the exact expressions in equations
(\ref{eqexact1}) and (\ref{eqexact2}) produced a positive-definite
covariance matrix but was significantly more time-consuming.  We
therefore restricted our fits to the two GAMA auto-power spectra and
excluded the cross-power spectrum; the growth-rate error predicted by
the Fisher matrix is worsened by only $0.2\%$ for the GAMA survey
specifications.  We note that, as justified by the Fisher matrix
forecasts below, the cross-power spectrum does add significant
information for galaxy samples with lower number densities ($n < 3
\times 10^{-4} \, h^3$ Mpc$^{-3}$).

We instead used the measured cross-power spectrum to provide some
independent validation of the modelling assumptions.  As an example,
the right-hand column of Fig.\ \ref{figpkmeas} compares the
cross-power spectrum measurements to the model fitted to the two
auto-power spectra, finding satisfactory agreement as judged by the
values of the $\chi^2$ statistic.

\begin{figure*}
\begin{center}
\resizebox{13cm}{!}{\rotatebox{270}{\includegraphics{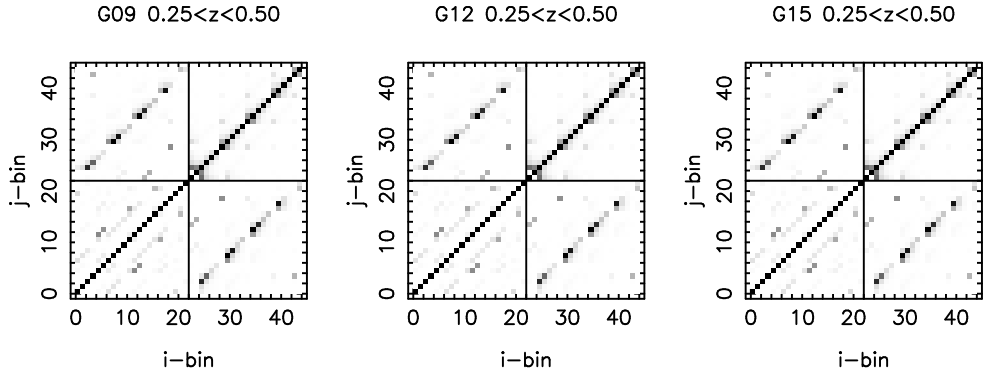}}}
\end{center}
\caption{The correlation matrices between the two auto-power spectra
  of the blue and red galaxy subsamples in the redshift interval $0.25
  < z < 0.5$, for the 3 GAMA regions.  The measurements are ordered by
  looping over the bins of $k_\perp$ and $k_\parallel$ (only plotting
  bins for which $|k| = \sqrt{k_\perp^2 + k_\parallel^2} < 0.3 \, h$
  Mpc$^{-1}$, constituting 22 bins) and then concatenating these
  results as $(P_1, P_2)$.  The result is a $44 \times 44$ covariance
  matrix containing a characteristic structure of diagonals, with
  strong correlations between different statistics measured in the
  same Fourier bins and weaker correlations between different Fourier
  bins.}
\label{figcov}
\end{figure*}

For model-fitting we defined a total data vector in which the
measurements of the two auto-power spectra were concatenated into a
longer vector $\hat{y}_i \equiv [\hat{P}_1(i), \hat{P}_2(i)]$ (for the
binned measurements) and $\hat{y}(\vec{k}) \equiv [\hat{P}_1(\vec{k}),
  \hat{P}_2(\vec{k})]$ (for the original Fourier modes).  Given that
the binned estimates of power are averages within each Fourier bin
$\hat{y}_i = (1/m_i) \sum_{\vec{k}} \hat{y}(\vec{k})$, where the sum
is over the $m_i$ Fourier modes $\vec{k}$ lying in bin $i$, then the
covariance of the binned estimates is
\begin{equation}
C_{ij} = \langle \delta \hat{y}_i \, \delta \hat{y}_j \rangle =
\frac{1}{m_i \, m_j} \sum_{\vec{k},\vec{k}'} \langle \hat{y}(\vec{k}) \,
\hat{y}(\vec{k}') \rangle
\label{eqallcov}
\end{equation}
We evaluated these covariance relations over the FFT grids for each
GAMA region, using equation (\ref{eqcov}).  Fig.\ \ref{figcov}
illustrates the structure of the resulting covariance matrices for the
$0.25 < z < 0.5$ auto-power spectrum measurements, with each displayed
as a correlation matrix $C_{ij}/\sqrt{C_{ii} \, C_{jj}}$.  We note the
characteristic structure of diagonals, with strong correlations
between different statistics measured in the same Fourier bins, and
weaker correlations between different Fourier bins.

We tested our determination of the covariance matrix using a large
ensemble of lognormal realizations.  For each realization, two
(correlated) populations of galaxies were created by Poisson-sampling
the same underlying density field using the GAMA survey selection
functions.  The diagonal and off-diagonal amplitudes of the lognormal
and analytic covariance matrices were in good agreement, with the
numerical values of the matrix elements differing by less than $10\%$.

\section{Model fits}
\label{secmodel}

\subsection{RSD modelling}

We fit the power spectrum measurements in each GAMA region using a
standard model for the redshift-space power spectrum as a function of
the cosine of the angle of the Fourier wavevector to the
line-of-sight, $\mu$:
\begin{align}
P_\alpha(k,\mu) &= \left[ b_\alpha^2 P_{\delta\delta}(k) + 2 b_\alpha f \mu^2 P_{\delta\theta}(k) + f^2 \mu^4 \, P_{\theta\theta}(k) \right] \nonumber \\
&\times e^{-k^2 \mu^2 \sigma_v^2 / H_0^2}
\label{eqpkmod1}
\end{align}
(Scoccimarro 2004) where, in terms of the divergence of the peculiar
velocity field $\theta$, $P_{\delta\delta}(k)$, $P_{\delta\theta}(k)$
and $P_{\theta\theta}(k)$ are the isotropic density-density,
density-$\theta$ and $\theta$-$\theta$ power spectra.  This model
combines the large-scale `Kaiser limit' amplitude correction with a
heuristic damping of power on smaller scales that describes a
leading-order perturbation theory correction.  Here, the free
parameter $\sigma_v$ has units of km s$^{-1}$ and $H_0 = 100 \, h$ km
s$^{-1}$ Mpc$^{-1}$.  When fitting multiple tracers, we make the
approximation that all populations of galaxies trace the same value of
$\sigma_v$ on large scales, as predicted by linear theory:
\begin{equation}
\sigma_v^2 = \frac{f^2 H_0^2}{6 \pi^2} \int P_{\theta\theta}(k) \, dk
\label{eqsigv}
\end{equation}
although, as stated above, we treat $\sigma_v$ as a free parameter to
allow for non-linearities in the matter clustering.  On large scales,
we neglect the contribution to equation (\ref{eqsigv}) from virialized
galaxy motions within dark matter halos.  Approximating
$P_{\theta\theta}$ as a linear power spectrum, the prediction of
equation (\ref{eqsigv}) in our fiducial cosmology is $\sigma_v = 334$
km s$^{-1}$.

We generated the matter power spectrum $P_{\delta\delta}$ in equation
(\ref{eqpkmod1}) using the `halofit' model (Smith et al.\ 2003) as
implemented by the {\tt CAMB} software package (Lewis, Challinor \&
Lasenby 2000) with the cosmological parameters fixed at values
inspired by fits to the CMB fluctuations measured by WMAP (Komatsu et
al.\ 2011): matter density $\Omega_{\rm m} = 0.27$, Hubble parameter
$h = 0.719$, spectral index $n_{\rm s} = 0.963$, baryon fraction
$\Omega_{\rm b}/\Omega_{\rm m} = 0.166$ and normalization $\sigma_8 =
0.8$.  We considered two different choices for producing the model
velocity power spectra $P_{\delta\theta}$ and $P_{\theta\theta}$.  In
our fiducial model, we used the large-scale limits of the velocity
power spectra $P_{\delta \theta} = P_{\theta \theta} = P_{\delta
  \delta}$, such that the model of equation (\ref{eqpkmod1})
simplified to:
\begin{equation}
P_\alpha(k,\mu) = P_{\delta\delta}(k) \, \left( b_\alpha + f \mu^2
\right)^2 \, e^{-k^2 \mu^2 \sigma_v^2 / H_0^2}
\label{eqpkmod2}
\end{equation}
Secondly, we investigated whether our results changed significantly if
we used the fitting formulae for $P_{\delta\theta}$ and
$P_{\theta\theta}$ in terms of $P_{\delta\delta}$, calibrated by
N-body simulations, proposed by Jennings et al.\ (2011).

Our model is hence characterized by four parameters $(f, b_1, b_2,
\sigma_v)$.  Given that $P_{\delta\delta}(k) \propto \sigma_8^2$,
where $\sigma_8$ characterizes the root-mean-square fluctuation of the
matter density in spheres of radius $8 \, h^{-1}$ Mpc, this parameter
set may also be written as $(f \sigma_8, b_1 \sigma_8, b_2 \sigma_8,
\sigma_v)$.  We compared the fits of the 4-parameter model to the
multiple tracers with fits of a 3-parameter model $(f, b_\alpha,
\sigma_v)$ to each individual galaxy subsample.  We performed the fits
by evaluating the $\chi^2$ statistic of each model for each survey
region using the full covariance matrix of equation (\ref{eqallcov}):
\begin{equation}
\chi^2 = \sum_{ij} (\hat{y}_i - y_{{\rm mod},i}) \, \left[ C^{-1}
  \right]_{ij} \, (\hat{y}_j - y_{{\rm mod},j})
\end{equation}
where in accordance with the notation of equation (\ref{eqallcov}),
$\hat{y}_i$ is a total data vector, concatenating the auto-power
spectra of the two subsamples, and $y_{{\rm mod},i}$ is the
corresponding model vector.  We assumed the measurements in each
survey region were independent, hence sum the values of $\chi^2$
corresponding to each model.

We fit the RSD model of equation (\ref{eqpkmod1}) to our measurements
in the range $k = \sqrt{k_\perp^2 + k_\parallel^2} < 0.3 \, h$
Mpc$^{-1}$ (noting that $\mu = k_\parallel/k$).  We fixed the
background cosmic expansion model and just varied the RSD parameters.
It is beyond the scope of this study to consider the Alcock-Paczynski
distortions that result from uncertainties in the cosmic
distance-scale, although we note in general that by improving
measurements of the growth rate, a multiple-tracer analysis also
enhances the determination of the geometrical Alcock-Paczynski
distortion, leading to improved distance and expansion measurements
for a galaxy sample.

We comment on the validity of the approximations of equations
(\ref{eqpkmod1}) and (\ref{eqpkmod2}).  First, in a similar analysis
of the WiggleZ Dark Energy Survey, Blake et al.\ (2011) established
that these models (including the free damping parameter) were an
acceptable approximation to a large suite of other approaches for
modelling non-linearities in RSD, including perturbation theory
techniques.  Indeed, both of these models ranked among the
best-performing models, as defined by the lowest values of $\chi^2$
and the stability of the fits when increasing the maximum wavenumber
fitted in the range $k_{\rm max} < 0.3 \, h$ Mpc$^{-1}$.

Secondly, in section \ref{secsim} below we demonstrate that these
techniques recovered the input growth rate (within an acceptable
margin of systematic error) in mock catalogues designed with similar
selection functions and galaxy bias factors as the GAMA populations.
We found that the model of equation (\ref{eqpkmod2}) produced no
detectable systematic bias in the growth rate, whereas the model of
equation (\ref{eqpkmod1}), using the Jennings et al.\ formulae,
resulted in an over-estimation of the growth rate in the simulation at
a level similar to the statistical error, justifying our decision to
choose equation (\ref{eqpkmod2}) as the fiducial model.  The results
of these sorts of tests depend on the clustering statistic and range
of scales being fitted (see also de la Torre \& Guzzo 2012).  At the
level of statistical precision of the GAMA measurement, conclusions
are unaffected by this choice of RSD model.

Thirdly, our parameter fits to the individual auto-power spectra alone
produce consistent values of $\sigma_v$ for each subsample, as
illustrated by Fig.\ \ref{figparfit} (for the purpose of comparison
with other $\sigma_v$ measurements in the literature, we note that
these values are 1-particle dispersions; a corresponding pairwise
dispersion would be larger by a factor of $\sqrt{2}$).  We explored
replacing the Gaussian damping term $e^{-k^2 \mu^2 \sigma_v^2 /
  H_0^2}$ by the Lorentzian $\left[ 1 + \left( k \mu \sigma_v / H_0
  \right)^2 \right]^{-1}$, finding a negligible difference in the
results.

\subsection{Parameter fits}

\begin{table*}
\begin{center}
\caption{Fits of RSD models in single-tracer and multiple-tracer
  analyses of subsamples of GAMA galaxies in two different redshift
  intervals $0 < z < 0.25$ and $0.25 < z < 0.5$, with effective
  redshifts $z=0.18$ and $0.38$, respectively.  Columns 3-6 display
  the results of fitting the 4-parameter model $(f,\sigma_v,b_1,b_2)$.
  Columns 7-8 are the fits of an alternative parameterization
  $(\beta_1,\sigma_v,b_1,b_2/b_1)$.  Column 9 provides the
  best-fitting values of $\chi^2$ and corresponding numbers of
  degrees-of-freedom.}
\label{tabparfit}
\begin{tabular}{ccccccccc}
\hline
Redshift & Sample & $f$ & $\sigma_v$ [km/s] & $b_1$ & $b_2$ & $\beta_1$ & $b_2/b_1$ & $\chi^2/{\rm dof}$ \\
\hline
$0.0<z<0.25$ & Blue & $0.49 \pm 0.14$ & $277 \pm  59$ & $0.891 \pm 0.038$ & - & $0.56 \pm 0.17$ & - & $ 41.6/ 63$ \\
& Red & $0.35 \pm 0.15$ & $246 \pm  57$ & - & $1.377 \pm 0.041$ & - & - & $ 53.9/ 63$ \\
& Joint & $0.49 \pm 0.12$ & $285 \pm  41$ & $0.894 \pm 0.038$ & $1.348 \pm 0.038$ & $0.57 \pm 0.15$ & $1.509 \pm 0.030$ & $163.1/128$ \\
\hline
$0.0<z<0.25$ & low-$L$ & $0.45 \pm 0.15$ & $267 \pm  59$ & $1.066 \pm 0.039$ & - & $0.43 \pm 0.15$ & - & $ 41.9/ 63$ \\
& high-$L$ & $0.35 \pm 0.15$ & $211 \pm  62$ & - & $1.480 \pm 0.043$ & - & - & $ 78.1/ 63$ \\
& Joint & $0.33 \pm 0.13$ & $192 \pm  65$ & $1.071 \pm 0.038$ & $1.467 \pm 0.039$ & $0.32 \pm 0.13$ & $1.371 \pm 0.020$ & $175.5/128$ \\
\hline
$0.25<z<0.5$ & Blue & $0.68 \pm 0.10$ & $269 \pm  34$ & $1.074 \pm 0.034$ & - & $0.64 \pm 0.10$ & - & $ 90.4/ 63$ \\
& Red & $0.48 \pm 0.11$ & $256 \pm  31$ & - & $1.707 \pm 0.035$ & - & - & $ 79.0/ 63$ \\
& Joint & $0.66 \pm 0.09$ & $286 \pm  23$ & $1.105 \pm 0.031$ & $1.664 \pm 0.030$ & $0.60 \pm 0.09$ & $1.508 \pm 0.027$ & $167.7/128$ \\
\hline
$0.25<z<0.5$ & low-$L$ & $0.63 \pm 0.09$ & $294 \pm  31$ & $1.283 \pm 0.020$ & - & $0.49 \pm 0.07$ & - & $ 75.6/ 63$ \\
& high-$L$ & $0.47 \pm 0.12$ & $224 \pm  37$ & - & $1.789 \pm 0.041$ & - & - & $ 75.7/ 63$ \\
& Joint & $0.57 \pm 0.08$ & $265 \pm  28$ & $1.283 \pm 0.020$ & $1.780 \pm 0.026$ & $0.45 \pm 0.07$ & $1.388 \pm 0.018$ & $147.5/128$ \\
\hline
\end{tabular}
\end{center}
\end{table*}

\begin{figure*}
\begin{center}
\resizebox{13cm}{!}{\rotatebox{270}{\includegraphics{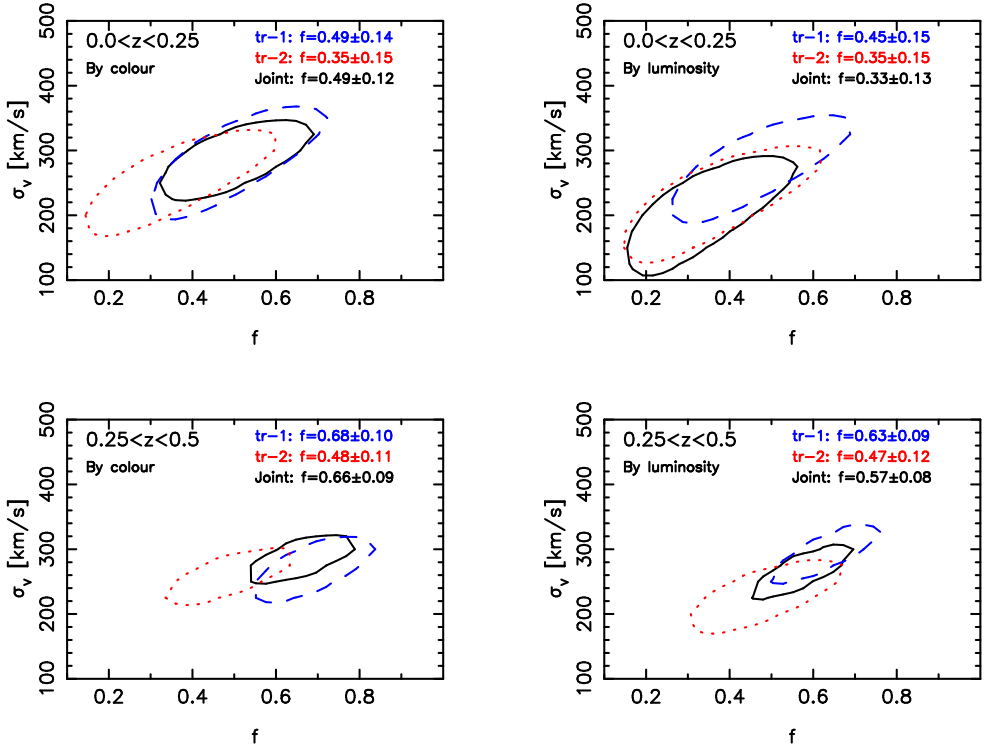}}}
\end{center}
\caption{Fits for the RSD parameters $(f,\sigma_v)$, marginalized over
  galaxy bias, for different redshift ranges and multiple-tracer
  subsamples (split by both colour and luminosity).  In each case we
  compare the fits to the individual subsamples (blue dashed and red
  dotted contours for the low-bias (`tr-1') and high-bias (`tr-2')
  sample, respectively) and the joint sample (black solid contours).
  The likelihood contours are all $68\%$ confidence regions.  The
  captions quote the 1D marginalized measurements of the growth rate.}
\label{figparfit}
\end{figure*}

In table \ref{tabparfit} we display the best-fitting parameters and
$68\%$ confidence regions (marginalized over all other parameters) for
various fits of these RSD models.  For each redshift range, defining
galaxy subsamples by either colour or luminosity, we compared fits of
the 3-parameter model $(f, \sigma_v, b)$ to each individual auto-power
spectrum with fits of the 4-parameter model $(f, \sigma_v, b_1, b_2)$
to the multiple-tracer auto-power spectra (with appropriate
covariance).  The results of these fits are shown in columns 3-6 of
the table.  The best-fitting values of $\chi^2$ (and corresponding
numbers of degrees of freedom) are listed in column 9; the model
produces a reasonable fit to the data.

We also considered the alternative parameterization $(\beta_1,
\sigma_v, b_1, b_2/b_1)$ where $\beta_1 = f/b_1$, to investigate
whether the multiple-tracer analysis allows the combinations of
parameters $f/b_1$ or $b_2/b_1$ to be determined with any additional
accuracy.  These results are shown in columns 7-8.  We found that the
ratio of the galaxy bias factors of the multiple populations,
$b_2/b_1$, was measured significantly more accurately for the
multiple-tracer fits than would be obtained by a naive propagation of
the errors in the individual bias factors in the single-tracer fits,
but the fractional errors in measuring $\beta_1$ were similar to those
in determining $f$.  We note that the precision afforded by a
multiple-tracer analysis for measuring bias ratios (which can be
carried out using the 1D monopole power spectra) could provide a
valuable test of models which predict the trend of bias with galaxy
luminosity or colour.

Fig.\ \ref{figparfit} shows likelihood contours in the space of $(f,
\sigma_v)$ marginalized over the bias parameter(s), comparing the
single-tracer and multiple-tracer fits.  In all cases we found that
the parameter measurements from different tracers were mutually
consistent, and that the fit to the combined data produced a
significant shrinkage in the size of the $68\%$ confidence region.  In
terms of the width of the $68\%$ confidence interval for the posterior
probability distribution of $f$, the multiple-tracer fits produced
reductions in the range $10$-$20\%$.  In section \ref{secfish} we will
demonstrate in a Fisher matrix analysis that truly large improvements
in the accuracy of determination of the growth rate require higher
galaxy number densities ($n > 10^{-2} \, h^3$ Mpc$^{-3}$).

\begin{figure}
\begin{center}
\resizebox{8cm}{!}{\rotatebox{270}{\includegraphics{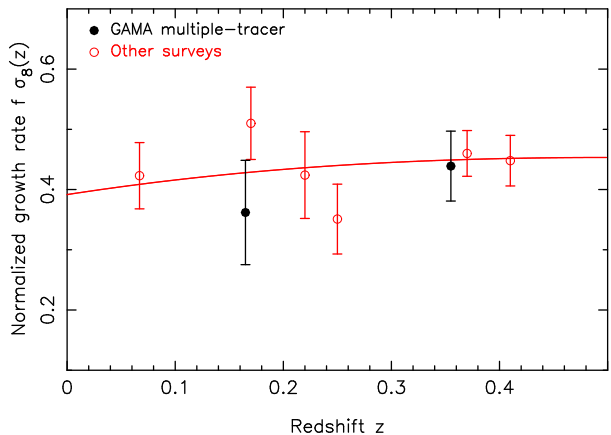}}}
\end{center}
\caption{Marginalized measurements of the normalized growth rate $f
  \sigma_8(z)$ fit to multiple-tracer GAMA galaxy subsamples split by
  colour.  The prediction of a flat $\Lambda$CDM model with matter
  density $\Omega_{\rm m} = 0.27$ and normalization $\sigma_8 = 0.8$
  is also shown as the solid line.  The open squares display the
  results of RSD analyses of a series of other galaxy surveys in a
  similar redshift range, taken from 6dFGS ($z=0.067$, Beutler et
  al.\ 2012), 2dFGRS ($z=0.17$, Hawkins et al.\ 2003), the SDSS
  Luminous Red Galaxy sample ($z=0.25$ and $z=0.37$, Samushia et
  al.\ 2012) and the WiggleZ Survey ($z=0.22$ and $z=0.41$, Blake et
  al.\ 2011).}
\label{figgrowth}
\end{figure}

Fig.\ \ref{figgrowth} displays the marginalized measurements of the
normalized growth rate $f \sigma_8(z)$ for the GAMA multiple-tracer
analysis split by colour, compared to the prediction of a flat
$\Lambda$CDM model with matter density $\Omega_{\rm m} = 0.27$ and
normalization $\sigma_8 = 0.8$.  The measurements of $f \sigma_8(z)$
in redshift slices $(0 < z < 0.25, 0.25 < z < 0.5)$ are $(0.36 \pm
0.09, 0.44 \pm 0.06)$, respectively.  We compared the GAMA
measurements with the published RSD analyses of a series of other
galaxy surveys in a similar redshift range, which are plotted as the
open squares in Fig.\ \ref{figgrowth}.  These measurements were taken
from 6dFGS ($z=0.067$, Beutler et al.\ 2012), 2dFGRS ($z=0.17$,
Hawkins et al.\ 2003), the SDSS Luminous Red Galaxy sample ($z=0.25$
and $z=0.37$, Samushia et al.\ 2012) and the WiggleZ Survey ($z=0.22$
and $z=0.41$, Blake et al.\ 2011).  Our GAMA measurements are
consistent with the results of these other surveys at similar
redshifts.

\section{Validation using N-body simulations}
\label{secsim}

We tested the validity of the non-linear RSD model of equation
(\ref{eqpkmod1}), in particular the amplitude of any systematic
modelling error that may impact the growth-rate measurements, by
fitting it to power spectrum measurements of dark matter halo
catalogues generated from N-body simulations.  We carried out these
tests using the GiggleZ N-body simulation (Poole et al.\ in
preparation), a $2160^3$ particle dark matter simulation run in a $1
\, h^{-1}$ Gpc box (with resulting particle mass $7.5 \times 10^9 \,
h^{-1} M_\odot$).  Bound structures were identified using {\tt
  Subfind} (Springel et al.\ 2001), which uses a friends-of-friends
(FoF) scheme followed by a sub-structure analysis to identify bound
overdensities within each FoF halo.  We employed each halo's maximum
circular velocity $V_{\rm max}$ as a proxy for mass, and used the
centre-of-mass velocities for each halo when introducing
redshift-space distortions.

We divided the GiggleZ simulation into 8 non-overlapping realizations
of the GAMA survey for the redshift range $0.25 < z < 0.5$, where each
realization consists of the 3 survey regions.  (We note that since we
are just using one simulation there will be low-level correlations
between these realizations deriving from common large-scale modes,
hence the scatter in results between the realizations may be slightly
under-estimated).  In each region we selected two populations of halos
which approximately reproduce the bias factors of the blue and red
GAMA populations, a `low-bias' set with $80 < V_{\rm max} < 135$ km
s$^{-1}$ and a `high-bias' set with $135 < V_{\rm max} < 999$ km
s$^{-1}$, and subsampled these halos using the full survey selection
functions.  We note that our intention here was not to produce full
mock GAMA catalogues, since we incorporated no information about
colour, luminosity or halo occupation distribution, but rather to
validate that the RSD model of equation (\ref{eqpkmod1}) was able to
reproduce the input growth rate of the N-body simulation on
quasi-linear scales, with minimal systematic error.

\begin{figure}
\begin{center}
\resizebox{8cm}{!}{\rotatebox{270}{\includegraphics{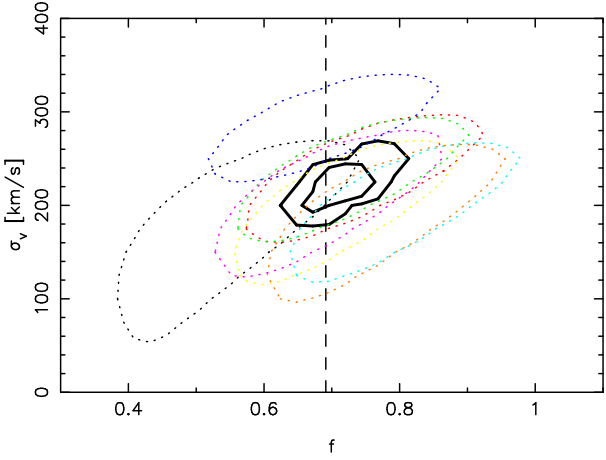}}}
\end{center}
\caption{Growth-rate fits to multiple-tracer power spectra measured in
  8 different realizations of the GAMA survey extracted from a large
  N-body simulation.  In each realization, two halo catalogues were
  extracted with bias factors close to two GAMA populations, and were
  subsampled in three survey regions using the appropriate selection
  functions for the $0.25 < z < 0.5$ redshift range.  The 8 sets of
  dotted coloured contours represent the $68\%$ confidence region of
  $(f,\sigma_v)$ fits (marginalized over bias parameters) to each of
  the 8 realizations, using the same RSD model and fitting range as
  applied to the GAMA data.  The solid black contours are $68\%$ and
  $95\%$ confidence regions obtained by combining these 8
  measurements, assuming that they were independent.  The vertical
  dotted line is the growth rate deduced from the input cosmological
  parameters of the simulation.}
\label{figmock}
\end{figure}

We measured the auto-power spectra of the two populations in each
survey region for each realization and fitted the RSD model of
equation (\ref{eqpkmod2}), using the same techniques we applied when
analyzing the real data (using the range $k < 0.3 \, h$ Mpc$^{-1}$).
Fig.\ \ref{figmock} shows the marginalized measurements of $(f,
\sigma_v)$ for each of the 8 realizations, with the $68\%$ confidence
region displayed as the dotted (coloured) lines.  The solid black
contours are the $68\%$ and $95\%$ confidence regions obtained by
combining these 8 measurements, assuming that they were independent.
The vertical dashed line indicates the predicted growth rate $f =
0.69$ based on the input cosmological parameters of the N-body
simulation (at $z=0.408$); the fits reveal no evidence for systematic
modelling errors.  The average best-fitting $\chi^2$ for the 8
realizations is $91.4$ for 128 degrees of freedom.

With the caveat that we only used 8 realizations, we compared the
errors in the measured growth rates of the simulations and data.  The
average error in the growth rate in the fits to the mock catalogues
was $\Delta f = 0.11$, compared to $\Delta f = 0.09$ for the data, and
the standard deviation in the best-fitting values for each realization
was $\sigma_f = 0.07$.  Given that these mock catalogues do not match
the galaxy populations of the data sample exactly, we consider the
more conservative value obtained from the data covariance matrix to be
the more reliable estimate.

\section{Fisher matrix forecasts}
\label{secfish}

We compared our measurements with Fisher matrix forecasts, which also
indicate how our results would extend to surveys with a different
design (also see McDonald \& Seljak 2009, White et al.\ 2009, Abramo
2012).  In this section we adopt the notation $P_{ij}$ to describe the
auto-power spectra between tracers (with $j = i$) and cross-power
spectra (with $j \ne i$).  We assume that the covariance matrix for
the measurement of $(P_{11}, P_{22}, P_{12})$ using an individual
Fourier mode $\vec{k} = (k,\mu)$ can be written following equation
(\ref{eqcovsimp}) as
\begin{equation}
C(\vec{k}) = \left( \begin{array}{ccc}
Q_1^2 & P_1 P_2 & Q_1 \sqrt{P_1 P_2} \\
P_1 P_2 & Q_2^2 & Q_2 \sqrt{P_1 P_2} \\
Q_1 \sqrt{P_1 P_2} & Q_2 \sqrt{P_1 P_2} & \frac{1}{2} (P_1 P_2 + Q_1 Q_2)
\end{array} \right)
\end{equation}
where we have written $P_i = P_{ii}$ and $Q_i = P_i + 1/n_i$, where
$n_i$ is the number density of the tracers.  The RSD power spectrum
model (using equation (\ref{eqpkmod2}) for simplicity) is then
\begin{equation}
P_{ij}(k,\mu) = (b_i + f \mu^2) \, (b_j + f \mu^2) \, P_m(k) \, e^{-
  k^2 \mu^2 \sigma_v^2 / H_0^2}
\end{equation}
where $b_i$ are the bias factors of the tracers.  The derivatives with
respect to the parameters are
\begin{align}
\frac{\partial P_{ij}}{\partial f} &= \left[ (b_i + b_j) \mu^2 + 2 f \mu^4 \right] P_m(k) \, e^{- k^2 \mu^2 \sigma_v^2/H_0^2} \nonumber \\
\frac{\partial P_{ii}}{\partial b_i} &= \frac{2 P_{ii}}{b_i + f \mu^2} \nonumber \\
\frac{\partial P_{ii}}{\partial b_j} &= 0 \hspace{5mm} (j \ne i) \nonumber \\
\frac{\partial P_{ij}}{\partial b_i} &= \frac{P_{ij}}{b_i + f \mu^2} \hspace{5mm} (j \ne i) \nonumber \\
\frac{\partial P_{ij}}{\partial \sigma_v^2} &= - \frac{k^2 \mu^2}{H_0^2} P_{ij}
\end{align}
The Fisher matrix of the parameter vector $p_\alpha = (f, \sigma_v^2,
b_1, b_2)$ is written
\begin{equation}
F_{\alpha\beta} = \sum_{k,\mu} m(k,\mu) \sum_{i,j} \frac{\partial
  P_{ij}(k,\mu)}{\partial p_\alpha} \, \left[ C(k,\mu)^{-1}
  \right]_{ij} \, \frac{\partial P_{ij}(k,\mu)}{\partial p_\beta}
\end{equation}
where $m(k,\mu)$ is the number of modes in a $(k,\mu)$ bin of width
$(\Delta k,\Delta \mu)$, which we deduce from the survey volume $V$ as
\begin{equation}
m(k,\mu) = \frac{V}{(2\pi)^3} \, 2 \pi \, k^2 \, \Delta k \, \Delta \mu
\end{equation}
We considered 5 bins in $\mu$ in the range $0 < \mu < 1$ and 6 bins in
$k$ in the range $0 < k < 0.3 \, h$ Mpc$^{-1}$, although our results
were not sensitive to the bin widths.  The covariance matrix of the
parameters follows as $C_{\alpha\beta} = (F^{-1})_{\alpha\beta}$, and
we focused in particular on the forecast error in the growth rate
measurement, $\Delta f = \sqrt{C_{11}} = (F^{-1})_{11}$.

In our fiducial model of the GAMA II survey we fixed the RSD
parameters $(f, \sigma_v) = (0.59, 300)$, number densities $n_i = 5
\times 10^{-3} \, h^3$ Mpc$^{-3}$, bias factors $(b_1, b_2) = (1.0,
1.4)$ and volume $V = 6.42 \times 10^6 \, h^{-3}$ Mpc$^3$.  These
values are representative of the two-sample dataset for $0 < z <
0.25$.  The forecast marginalized error in the growth rate for this
case is $\Delta f = 0.096$ for the multiple-tracer fits, and $\Delta f
= 0.124$ and $0.156$ for the low-bias and high-bias single-tracer
fits, respectively (such that the multiple-tracer analysis produces a
$\approx 20$ per cent improvement compared to the low-bias case).
These forecasts are a little better than, although comparable to, the
measurements quoted in table \ref{tabparfit}, and we note that the
Fisher matrix forecast assumes a perfect-cuboid survey with no
correlations between different Fourier modes.

\begin{figure}
\begin{center}
\resizebox{8cm}{!}{\rotatebox{0}{\includegraphics{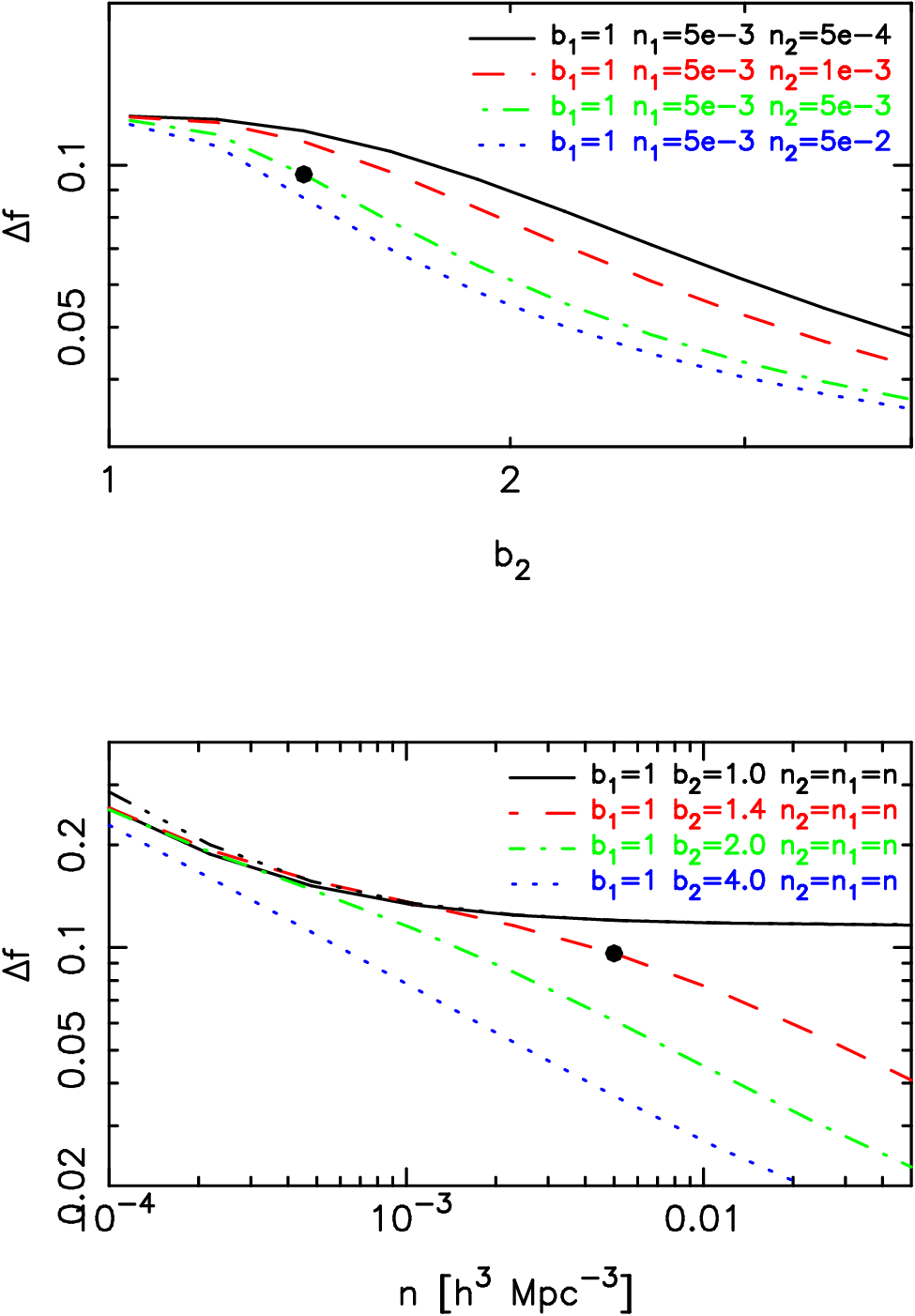}}}
\end{center}
\caption{Fisher matrix forecasts for the error in the growth rate,
  $\Delta f$, marginalized over the other RSD parameters.  We consider
  two-tracer survey configurations varying the bias parameters $(b_1,
  b_2)$ and number densities $(n_1, n_2)$, fixing the survey volume $V
  = 6.42 \times 10^6 \, h^{-3}$ Mpc$^3$ and $b_1 = 1$ for all cases.
  In the upper panel we fix $n_1 = 5 \times 10^{-3} \, h^3$ Mpc$^{-3}$
  and plot $\Delta f$ as a function of $b_2$ for various choices of
  $n_2$.  In the lower panel we plot $\Delta f$ as a function of $n =
  n_1 = n_2$ for various choices of $b_2$.  The dash-triple-dotted
  black curve in the lower panel, compared to the solid black curve,
  shows the effect of dropping the cross-power spectrum information.
  The solid circles in the panels indicate the fiducial GAMA values of
  these parameters.  Changing the survey volume $V$ will simply scale
  the results by $\Delta f \propto V^{-1/2}$.}
\label{figfishpred}
\end{figure}

We then considered two sets of variations which allow us to explore
other survey designs:
\begin{itemize}
\item Varying the bias factor of the second tracer in the range $1 <
  b_2 < 4$ for different choices of $n_2$, fixing $b_1 = 1$ and $n_1
  = 5 \times 10^{-3} \, h^3$ Mpc$^{-3}$.
\item Varying the number density of both tracers in the range $1
  \times 10^{-4} < n_i < 5 \times 10^{-2} \, h^3$ Mpc$^{-3}$ for
  different choices of $b_2$, fixing $b_1 = 1$.
\end{itemize}
The results are displayed in Fig.\ \ref{figfishpred}, with the solid
circles indicating the fiducial GAMA case quoted above.

The upper panel of Fig.\ \ref{figfishpred} indicates the improvement
in the multiple-tracer growth rate measurement that results as the
difference between the bias factors of the galaxy populations
increases.  For $n_2 = n_1 = 5 \times 10^{-3} \, h^3$ Mpc$^{-3}$ and
$b_1 = 1$, the growth rate measurement improves by $(8, 22, 35, 44,
51)\%$ for $b_2 = (1.2, 1.4, 1.6, 1.8, 2.0)$.  These forecast gains
will be impacted by the practical difficulty of maintaining a high
target number density as galaxy bias increases, as described by the
set of lines for different values of $n_2$ in the upper panel of
Fig.\ \ref{figfishpred}.

The lower panel of Fig.\ \ref{figfishpred} displays the increasing
efficacy of the multiple-tracer method as the number density of the
galaxy populations increases.  For $n > 10^{-3} \, h^3$ Mpc$^{-3}$ the
gains from single-tracer RSD saturate (as indicated by the solid black
line); but the growth rate measurement from multiple-tracers improves
by $(12, 22, 37, 53, 66)\%$ for $n = (0.23, 0.5, 1.1, 2.4, 5.2) \times
10^{-2} \, h^3$ Mpc$^{-3}$ assuming $(b_1, b_2) = (1.0, 1.4)$.  The
black dash-triple-dotted line in Fig.\ \ref{figfishpred}, which should
be compared with the black solid line, illustrates the effect of
dropping the information from the cross-power spectrum.  For low
values of number density $n < 10^{-3} \, h^3$ Mpc$^{-3}$ the
cross-power spectrum adds some information due to shot noise.  For
high number density $n > 10^{-3} \, h^3$ Mpc$^{-3}$ the cross-power
spectrum may be entirely predicted from the two auto-power spectrum
(under the assumption of linear galaxy bias) and hence its inclusion
does not improve the growth-rate measurements within the assumed RSD
model.

\section{Summary}
\label{secconc}

In this study we have presented the first observational
multiple-tracer analysis of redshift-space distortions using data from
the Galaxy and Mass Assembly survey.  We performed a Fourier analysis
of the two auto-power spectra of galaxy populations split by both
colour and luminosity, deriving new expressions for the covariances
between these measurements in terms of a general survey selection
function and weighting scheme, and verified our results by also
measuring the cross-power spectrum.  We fit models to the
redshift-space power spectra in terms of the gravitational growth
rate, $f$, linear galaxy bias factors and an empirical non-linear
damping parameter.  We find that, in the case of GAMA, the
multiple-tracer analysis produces an improvement in the measurement
accuracy of $f$ by $10$-$20\%$ (depending on the sample).  The growth
rates determined from the separate populations, split by colour and
luminosity, are consistent, showing no evidence for strong systematic
modelling errors.  The precision of our measurements is similar to a
Fisher matrix forecast, which indicates how our analyses would extend
to surveys with a different design: for samples with higher number
densities or bias factor differentials, much stronger improvements in
the accuracy of growth rate determination are expected.  We tested our
methodology using mock catalogues from N-body simulations,
demonstrating that the systematic error in the measured growth rate
was much smaller than the statistical error.  The normalized
gravitational growth rate determined in two independent redshift
slices, $f \sigma_8(z=0.18) = 0.36 \pm 0.09$ and $f \sigma_8(z=0.38) =
0.44 \pm 0.06$ using multiple-tracer subsamples selected by colour, is
consistent with results from other RSD surveys in a similar redshift
range, and with standard $\Lambda$ Cold Dark Matter models.

\section*{Acknowledgments}

CB acknowledges the support of the Australian Research Council through
the award of a Future Fellowship, and is grateful for useful feedback
from Yan-Chuan Cai and Felipe Marin.

GAMA is a joint European-Australasian project based around a
spectroscopic campaign using the Anglo-Australian Telescope.  The GAMA
input catalogue is based on data taken from the Sloan Digital Sky
Survey and the UKIRT Infrared Deep Sky Survey.  Complementary imaging
of the GAMA regions is being obtained by a number of independent
survey programs including GALEX MIS, VST KiDS, VISTA VIKING, WISE,
Herschel-ATLAS, GMRT and ASKAP providing UV to radio coverage.  GAMA
is funded by the STFC (UK), the ARC (Australia), the AAO, and the
participating institutions.  The GAMA website is {\tt
  http://www.gama-survey.org/}.

\appendix

\section{Derivation of auto-power and cross-power spectrum estimators and covariances}

\subsection{Fourier conventions}

First we note our conventions for Fourier transforms and inverse
Fourier transforms:
\begin{equation}
{\rm FT}(y) = \tilde{y}(\vec{k}) = \frac{1}{V} \int y(\vec{x}) \,
e^{i\vec{k}.\vec{x}} \, d^3\vec{x}
\end{equation}
\begin{equation}
{\rm IFT}(\tilde{y}) = y(\vec{x}) = \frac{V}{(2\pi)^3} \int
\tilde{y}(\vec{k}) \, e^{-i\vec{k}.\vec{x}} \, d^3\vec{k}
\end{equation}
where $V$ is the Fourier volume.  Some useful relations involving the
Dirac delta-function $\delta_D$ and its transform are:
\begin{equation}
\int e^{i\vec{k}.\vec{x}} \, d^3\vec{x} = V \, \tilde{\delta}_D(\vec{k})
\end{equation}
\begin{equation}
\int e^{i\vec{k}.\vec{x}} \, d^3\vec{k} = \frac{(2\pi)^3}{V} \,
\delta_D(\vec{x})
\end{equation}
\begin{equation}
\int \delta_D(\vec{x}-\vec{x}_0) \, \delta^3\vec{x} = V
\end{equation}
\begin{equation}
\int y(\vec{x}) \, \delta_D(\vec{x}-\vec{x}_0) \, \delta^3\vec{x} = V
\, y(\vec{x}_0)
\end{equation}
\begin{equation}
\int \tilde{\delta}_D(\vec{k}-\vec{k}_0) \, \delta^3\vec{k} =
\frac{(2\pi)^3}{V}
\end{equation}
\begin{equation}
\int \tilde{y}(\vec{k}) \, \tilde{\delta}_D(\vec{k}-\vec{k}_0) \,
\delta^3\vec{k} = \frac{(2\pi)^3}{V} \, \tilde{y}(\vec{k}_0)
\end{equation}
It is also useful to list our conventions for evaluating Fast Fourier
Transforms, which we will employ in practice for implementing these
calculations:
\begin{equation}
{\rm FFT}(y) = \sum_{\vec{x}_i} y(\vec{x}_i) \, e^{i\vec{k}.\vec{x}_i}
\end{equation}
\begin{equation}
{\rm IFFT}(\tilde{y}) = \sum_{\vec{k}_i} \tilde{y}(\vec{k}_i) \,
e^{-i\vec{k}_i.\vec{x}}
\end{equation}
Noting the equivalences $(1/V) \int d^3\vec{x} \equiv (1/N_c)
\sum_{\vec{x}}$ and $[V/(2\pi)^3] \int d^3\vec{k} \equiv
\sum_{\vec{k}}$, we deduce that ${\rm FFT}(y) = N_c \, {\rm FT}(y)$
and ${\rm IFFT}(\tilde{y}) = {\rm IFT}(\tilde{y})$, where $N_c$ is the
total number of FFT cells.

\subsection{Estimator for auto-power spectrum}

We first develop the estimator for the auto-power spectrum of a galaxy
number density distribution $n(\vec{x})$, given an underlying
selection function $\langle n(\vec{x}) \rangle$ describing the average
over many realizations, and allowing for a general weighting function
$w(\vec{x})$.  This derivation follows Feldman, Kaiser \& Peacock
(1994) and Smith (2009); we will then provide the extension to the
galaxy cross-power spectrum and the various covariances.  The
normalization of the number density in terms of the total number of
galaxies $N$ is such that
\begin{equation}
\int n(\vec{x}) \, d^3\vec{x} = N
\end{equation}
First we define the weighted galaxy overdensity
\begin{equation}
\delta(\vec{x}) = w(\vec{x}) \left[ n(\vec{x}) - \langle n(\vec{x})
\rangle \right]
\end{equation}
and consider the Fourier transform of this expression,
$\tilde{\delta}(\vec{k})$.  In order to perform this evaluation it is
convenient to split the sample volume into many small cells $i$ at
positions $\vec{x}_i$ with infinitesimal volumes $\delta V_i$, such
that the number of galaxies $N_i$ in the $i^{\rm th}$ cell is 0 or 1,
and we can write the number density distribution as
\begin{equation}
n(\vec{x}) = \frac{1}{V} \sum_i N_i \, \delta_D(\vec{x}-\vec{x}_i)
\end{equation}
which satisfies $\int n(\vec{x}) \, d^3\vec{x} = \sum_i N_i = N$.
Writing the weighted number density $n_w(\vec{x}) \equiv w(\vec{x}) \,
n(\vec{x})$, we find that
\begin{equation}
\tilde{n}_w(\vec{k}) = \frac{1}{V} \int n_w(\vec{x}) \, e^{i
  \vec{k}.\vec{x}} \, d^3\vec{x} = \frac{1}{V} \sum_i w_i \, N_i \,
e^{i \vec{k}.\vec{x}_i}
\end{equation}
hence
\begin{equation}
\tilde{\delta}(\vec{k}) = \tilde{n}_w(\vec{k}) - \langle
\tilde{n}_w(\vec{k}) \rangle = \frac{1}{V} \sum_i w_i \left( N_i -
\langle N_i \rangle \right) e^{i \vec{k}.\vec{x}_i}
\end{equation}
Then, using the fact that
\begin{equation}
\langle \left( N_i - \langle N_i \rangle \right) \left( N_j - \langle
N_j \rangle \right) \rangle = \langle N_i \, N_j \rangle - \langle N_i
\rangle \langle N_j \rangle
\end{equation}
we find that
\begin{align}
&\langle \tilde{\delta}(\vec{k}) \, \tilde{\delta}^*(\vec{k}') \rangle = \nonumber \\
&\frac{1}{V^2} \sum_{i,j} w_i \, w_j \left( \langle N_i \, N_j \rangle - \langle N_i \rangle \langle N_j \rangle \right) e^{i (\vec{k}.\vec{x}_i - \vec{k}'.\vec{x}_j)}
\label{eqautosum}
\end{align}
We evaluate this double sum by splitting it into two parts, with $j=i$
and $j \ne i$.  The part of the sum with $j=i$ can be simplified using
\begin{equation}
\langle N_i^2 \rangle - \langle N_i \rangle^2 = \langle N_i \rangle =
\langle n_i \rangle \, \delta V_i
\end{equation}
which holds given that $N_i^2 = N_i$ (for $N_i = 0$ or $1$) and
$\langle N_i \rangle^2 \propto (\delta V_i)^2$ is negligible.  We can
express the part of the sum with $j \ne i$ in terms of the galaxy
correlation function $\xi$ using
\begin{equation}
\langle N_i \, N_j \rangle - \langle N_i \rangle \langle N_j \rangle =
(\langle n_i \rangle \delta V_i) \, (\langle n_j \rangle \delta V_j)
\, \xi(\vec{x}_i, \vec{x}_j)
\label{eqxi}
\end{equation}
Making these substitutions,
\begin{align}
&\langle \tilde{\delta}(\vec{k}) \, \tilde{\delta}^*(\vec{k}') \rangle = \nonumber \\
&\frac{1}{V^2} \sum_{i \ne j} w_i \, w_j \, \langle n_i \rangle \, \langle n_j \rangle \, \delta V_i \, \delta V_j \, \xi(\vec{x}_i, \vec{x}_j) \, e^{i (\vec{k}.\vec{x}_i - \vec{k}'.\vec{x}_j)} \nonumber \\
&+ \frac{1}{V^2} \sum_i w_i^2 \, \langle n_i \rangle \, \delta V_i \, e^{i(\vec{k}-\vec{k}').\vec{x}_i}
\end{align}
Now we transform the sums into integrals and substitute the relation
\begin{equation}
\xi(\vec{x}, \vec{x}') = \frac{1}{(2\pi)^3} \int P(\vec{k}'') \, e^{-i
  \vec{k}''.(\vec{x}-\vec{x}')} \, d^3\vec{k}''
\label{eqpk}
\end{equation}
between the correlation function and auto-power spectrum $P(\vec{k})$
in volume units.  After some algebra we find:
\begin{align}
&\langle \tilde{\delta}(\vec{k}) \, \tilde{\delta}^*(\vec{k}') \rangle = \nonumber \\
&\frac{1}{(2\pi)^3} \int P(\vec{k}'') \, \tilde{n}_w(\vec{k}-\vec{k}'') \, \tilde{n}_w^*(\vec{k}'-\vec{k}'') \, d^3\vec{k}'' \nonumber \\
&+ \frac{1}{V^2} \int w(\vec{x})^2 \, n(\vec{x}) \, e^{i(\vec{k}-\vec{k}').\vec{x}} \, d^3\vec{x}
\label{eqexact1}
\end{align}
where for clarity we have dropped the angled brackets in the symbols
$n$ and $n_w$ in this and all subsequent equations.  If $P(\vec{k})$
varies sufficiently slowly compared to the width of
$\tilde{n}_w(\delta \vec{k})$, we can approximate the first term as
\begin{equation}
\frac{1}{(2\pi)^3} \, P(\vec{k}) \int \tilde{n}_w (\vec{k}-\vec{k}'') \, \tilde{n}_w^*(\vec{k}'-\vec{k}'') \, d^3\vec{k}''
\label{eqapprox1}
\end{equation}
where writing $\tilde{n}_w(\vec{k}) = (1/V) \int n_w(\vec{x}) \,
e^{i\vec{k}.\vec{x}} \, d^3\vec{x}$ we find that
\begin{align}
&\int \tilde{n}_w(\vec{k}-\vec{k}'') \, \tilde{n}_w^*(\vec{k}'-\vec{k}'') \, d^3\vec{k}'' = \nonumber \\
&\frac{(2\pi)^3}{V^2} \int n_w^2(\vec{x}) \, e^{i(\vec{k}-\vec{k}').\vec{x}} \, d^3\vec{x}
\end{align}
Hence we derive the final expression
\begin{equation}
\langle \tilde{\delta}(\vec{k}) \, \tilde{\delta}^*(\vec{k}') \rangle
\approx \frac{1}{V} \left[ P(\vec{k}) \, \tilde{Q}(\vec{k}-\vec{k}') +
  \tilde{S}(\vec{k}-\vec{k}') \right]
\label{eqd1d1}
\end{equation}
in terms of
\begin{align}
Q(\vec{x}) &= n_w^2(\vec{x}) \equiv n^2(\vec{x}) \, w^2(\vec{x}) \\
S(\vec{x}) &= n(\vec{x}) \, w^2(\vec{x})
\end{align}
Considering the special case $\vec{k}' = \vec{k}$ we see that an
estimator for the auto-power spectrum is
\begin{equation}
\hat{P}(\vec{k}) = \frac{V \, |\tilde{\delta}(\vec{k})|^2 -
  \tilde{S}(0)}{\tilde{Q}(0)}
\end{equation}
such that
\begin{equation}
\langle \hat{P}(\vec{k}) \rangle = \frac{V \, \langle
  |\tilde{\delta}(\vec{k})|^2 \rangle - \tilde{S}(0)}{\tilde{Q}(0)}
\approx P(\vec{k})
\end{equation}
where we note that the exact expression is the convolution
\begin{equation}
\langle \hat{P}(\vec{k}) \rangle = \frac{V^3}{(2\pi)^3} \int
P(\vec{k}') \, |\tilde{n}_w(\vec{k}-\vec{k}')|^2 \, d^3\vec{k}'
\end{equation}
We note the special case of a constant selection function $n(\vec{x})
= n_0 = N/V$ and weights $w(\vec{x}) = 1$.  In this case, $Q =
N^2/V^2$ and $S = N/V$, such that
\begin{equation}
\hat{P}(\vec{k}) = V \left[ \frac{V^2 \, |\tilde{\delta}(\vec{k})|^2 -
    N}{N^2} \right]
\end{equation}
Converting these relations to an FFT-based estimator, we grid the
galaxy number distribution into the FFT cells and write this number
distribution as $N(\vec{x})$.  Imposing the normalization that
$\sum_{\vec{x}} N(\vec{x}) = N$, we find that $N(\vec{x}) = (V/N_c) \,
n(\vec{x})$.  We define the selection function grid as $W(\vec{x})$,
adopting the normalization convention that $\sum_{\vec{x}} W(\vec{x})
= 1$.  In this case, $W(\vec{x}) = (V/N_c N) \langle n(\vec{x})
\rangle$.  Writing $N_w(\vec{x}) = w(\vec{x}) \, N(\vec{x})$ and
$W_w(\vec{x}) = w(\vec{x}) \, W(\vec{x})$, we have
\begin{equation}
\tilde{\delta}(\vec{k}) = \frac{1}{V} \left[ \tilde{N}_w - N \,
  \tilde{W}_w \right]
\end{equation}
where $\tilde{N}_w \equiv {\rm FFT}(N_w)$ and $\tilde{W}_w \equiv {\rm
  FFT}(W_w)$.  The power spectrum estimator becomes
\begin{equation}
\hat{P}(\vec{k}) = V \, \left[ \frac{|\tilde{N}_w - N \,
    \tilde{W}_w|^2 - N \sum W(\vec{x}) w(\vec{x})^2}{N_c \, N^2 \sum
    W(\vec{x})^2 \, w(\vec{x})^2} \right]
\end{equation}

\subsection{Estimator for cross-power spectrum}

The development of the cross-power spectrum estimator follows a
similar course, where we consider the two galaxy overdensity fields
\begin{equation}
\delta_1(\vec{x}) = w_1(\vec{x}) \left[ n_1(\vec{x}) - \langle
  n_1(\vec{x}) \rangle \right]
\end{equation}
\begin{equation}
\delta_2(\vec{x}) = w_2(\vec{x}) \left[ n_2(\vec{x}) - \langle
  n_2(\vec{x}) \rangle \right]
\end{equation}
The generalization of equation (\ref{eqautosum}) is
\begin{align}
&\langle \tilde{\delta}_1(\vec{k}) \, \tilde{\delta}_2^*(\vec{k}')
\rangle = \nonumber \\
&\frac{1}{V^2} \sum_{i,j} w_{1,i} \, w_{2,j} \left( \langle N_{1,i} \, N_{2,j}
\rangle - \langle N_{1,i} \rangle \langle N_{2,j} \rangle \right) e^{i
  (\vec{k}.\vec{x}_i - \vec{k}'.\vec{x}_j)}
\end{align}
The terms with $j = i$ now vanish, and the equivalent of equations
(\ref{eqxi}) and (\ref{eqpk}) now involve the cross-correlation
function $\xi_c(\vec{x},\vec{x}')$ and cross-power spectrum
$P_c(\vec{k})$.  We obtain
\begin{align}
&\langle \tilde{\delta}_1(\vec{k}) \, \tilde{\delta}_2^*(\vec{k}')
\rangle = \nonumber \\
&\frac{1}{(2\pi)^3} \int P_c(\vec{k}'') \, \tilde{n}_{w,1}(\vec{k}-\vec{k}'') \, \tilde{n}_{w,2}^*(\vec{k}'-\vec{k}'') \, d^3\vec{k}''
\label{eqexact2}
\end{align}
with the approximation, which again holds if $P_c(\vec{k})$ varies
sufficiently slowly compared to the width of
$\tilde{n}_{w,\alpha}(\delta \vec{k})$,
\begin{equation}
\langle \tilde{\delta}_1(\vec{k}) \, \tilde{\delta}_2^*(\vec{k}')
\rangle \approx \frac{P_c(\vec{k})}{V^2} \int n_{w,1}(\vec{x}) \,
n_{w,2}(\vec{x}) \, e^{i(\vec{k}-\vec{k}').\vec{x}} \, d^3\vec{x}
\label{eqapprox2}
\end{equation}
We define
\begin{equation}
Q_c(\vec{x}) = n_{w,1}(\vec{x}) \, n_{w,2}(\vec{x}) = w_1(\vec{x}) \,
n_1(\vec{x}) \, w_2(\vec{x}) \, n_2(\vec{x})
\end{equation}
such that
\begin{equation}
\langle \tilde{\delta}_1(\vec{k}) \, \tilde{\delta}_2^*(\vec{k}')
\rangle \approx \frac{1}{V} \, P_c(\vec{k}) \,
\tilde{Q}_c(\vec{k}-\vec{k}')
\label{eqd1d2}
\end{equation}
The estimator for the cross-power spectrum is then written
\begin{align}
\hat{P}_c(\vec{k}) &= \frac{ V \, {\rm Re} \left\{ \tilde{\delta}_1(\vec{k}) \, \tilde{\delta}_2^*(\vec{k}) \right\} }{\tilde{Q}_c(0)} \nonumber \\
&= \frac{ V \, \left[ \tilde{\delta}_1(\vec{k}) \, \tilde{\delta}_2^*(\vec{k}) + \tilde{\delta}_1^*(\vec{k}) \, \tilde{\delta}_2(\vec{k}) \right]}{2 \, \tilde{Q}_c(0)}
\end{align}
such that it is symmetric in the two indices, and
\begin{equation}
\langle \hat{P}_c(\vec{k}) \rangle = \frac{V \, \left[ \langle
  \tilde{\delta}_1(\vec{k}) \, \tilde{\delta}_2^*(\vec{k}) \rangle +
  \langle \tilde{\delta}_1^*(\vec{k}) \, \tilde{\delta}_2(\vec{k})
  \rangle \right]}{2 \, \tilde{Q}_c(0)} \approx P_c(\vec{k})
\end{equation}
with the exact expression
\begin{equation}
\langle \hat{P}_c(\vec{k}) \rangle = \frac{V^3}{(2\pi)^3} \int
P_c(\vec{k}'') \, \tilde{n}_{w,1}(\vec{k}-\vec{k}'') \,
\tilde{n}_{w,2}^*(\vec{k}'-\vec{k}'') \, d^3\vec{k}''
\end{equation}
The FFT-based estimator is
\begin{equation}
\hat{P}_c(\vec{k}) = \frac{V {\rm Re} \{ [\tilde{N}_{w,1} - N_1 \,
    \tilde{W}_{w,1}] [\tilde{N}_{w,2} - N_2 \, \tilde{W}_{w,2}]^*
  \}}{N_c \, N_1 \, N_2 \, \sum W_1(\vec{x}) \, w_1(\vec{x}) \,
  W_2(\vec{x}) \, w_2(\vec{x})}
\end{equation}

\subsection{Covariance between estimators}

We now consider the covariance between the estimators for the
auto-power spectra of the two galaxy populations, $\hat{P}_1$ and
$\hat{P}_2$, and the estimator for the cross-power spectrum
$\hat{P}_c$.  Defining $\delta \hat{P}(\vec{k}) \equiv
\hat{P}(\vec{k}) - \langle \hat{P}(\vec{k}) \rangle$, these
covariances can be written
\begin{align}
\langle \delta \hat{P}_1(\vec{k}) \, \delta \hat{P}_1(\vec{k}') \rangle &= \frac{V^2 \, \langle \tilde{\delta}_1(\vec{k}) \, \tilde{\delta}_1^*(\vec{k}) \, \tilde{\delta}_1(\vec{k}') \, \tilde{\delta}_1^*(\vec{k}') \rangle}{\tilde{Q}_1(0)^2} \\
\langle \delta \hat{P}_1(\vec{k}) \, \delta \hat{P}_2(\vec{k}') \rangle &= \frac{V^2 \, \langle \tilde{\delta}_1(\vec{k}) \, \tilde{\delta}_1^*(\vec{k}) \, \tilde{\delta}_2(\vec{k}') \, \tilde{\delta}_2^*(\vec{k}') \rangle}{\tilde{Q}_1(0) \, \tilde{Q}_2(0)} \\
\langle \delta \hat{P}_1(\vec{k}) \, \delta \hat{P}_c(\vec{k}') \rangle &= \frac{V^2}{2 \, \tilde{Q}_1(0) \, \tilde{Q}_c(0)} \times \nonumber \\
[ &\langle \tilde{\delta}_1(\vec{k}) \, \tilde{\delta}_1^*(\vec{k}) \, \tilde{\delta}_1(\vec{k}') \, \tilde{\delta}_2^*(\vec{k}') \rangle \nonumber \\
+ &\langle \tilde{\delta}_1(\vec{k}) \, \tilde{\delta}_1^*(\vec{k}) \, \tilde{\delta}_1^*(\vec{k}') \, \tilde{\delta}_2(\vec{k}') \rangle ] \\
\langle \delta \hat{P}_c(\vec{k}) \, \delta \hat{P}_c(\vec{k}') \rangle &= \frac{V^2}{4 \, \tilde{Q}_c(0)^2} \times \nonumber \\
[ &\langle \tilde{\delta}_1(\vec{k}) \, \tilde{\delta}_2^*(\vec{k}) \, \tilde{\delta}_1(\vec{k}') \, \tilde{\delta}_2^*(\vec{k}') \rangle \nonumber \\
+ &\langle \tilde{\delta}_1(\vec{k}) \, \tilde{\delta}_2^*(\vec{k}) \, \tilde{\delta}_1^*(\vec{k}') \, \tilde{\delta}_2(\vec{k}') \rangle \nonumber \\
+ &\langle \tilde{\delta}_1^*(\vec{k}) \, \tilde{\delta}_2(\vec{k}) \, \tilde{\delta}_1(\vec{k}') \, \tilde{\delta}_2^*(\vec{k}') \rangle \nonumber \\
+ &\langle \tilde{\delta}_1^*(\vec{k}) \, \tilde{\delta}_2(\vec{k}) \, \tilde{\delta}_1^*(\vec{k}') \, \tilde{\delta}_2(\vec{k}') \rangle ]
\end{align}
Taking the first expression as an example, these relations may be
evaluated by substituting $\tilde{\delta}(\vec{k}) = \sum_{\vec{x}}
\delta(\vec{x}) \, e^{i \vec{k}.\vec{x}}$.  We can then write the
product $\langle \tilde{\delta}(\vec{k}) \, \tilde{\delta}^*(\vec{k})
\, \tilde{\delta}(\vec{k}') \, \tilde{\delta}^*(\vec{k}') \rangle$ as
\begin{equation}
\sum_{\vec{x}_1,\vec{x}_2,\vec{x}_3,\vec{x}_4} \langle
\delta(\vec{x}_1) \delta(\vec{x}_2) \delta(\vec{x}_3)
\delta(\vec{x}_4) \rangle \, e^{i[\vec{k}.(\vec{x}_1-\vec{x}_2) +
    \vec{k}'.(\vec{x}_3-\vec{x}_4)]}
\end{equation}
Expectation values of individual terms in the product satisfy $\langle
\delta(\vec{x}) \rangle = 0$; non-zero terms are those in which the
indices of the sum satisfy $(1=2,3=4)$, $(1=3,2=4)$ or $(1=4,2=3)$.
Splitting the sum into these combinations, it can be expressed as
\begin{align}
\sum_{\vec{x},\vec{x}'} \{ \, &\langle \delta(\vec{x}) \delta(\vec{x}) \rangle \langle \delta(\vec{x}') \delta(\vec{x}') \rangle \nonumber \\
+ \, &\langle \delta(\vec{x}) \delta(\vec{x}) \rangle \langle \delta(\vec{x}') \delta(\vec{x}') \rangle \, e^{i(\vec{k}-\vec{k}').(\vec{x}-\vec{x}')} \nonumber \\
+ \, &\langle \delta(\vec{x}) \delta(\vec{x}) \rangle \langle \delta(\vec{x}') \delta(\vec{x}') \rangle \, e^{i(\vec{k}+\vec{k}').(\vec{x}-\vec{x}')} \, \}
\end{align}
We obtain
\begin{align}
\langle \delta \hat{P}_1(\vec{k}) \, \delta \hat{P}_1(\vec{k}') \rangle &= \frac{V^2 \, | \langle \tilde{\delta}_1(\vec{k}) \, \tilde{\delta}_1^*(\vec{k}') \rangle |^2}{\tilde{Q}_1(0)^2} \\
\langle \delta \hat{P}_1(\vec{k}) \, \delta \hat{P}_2(\vec{k}') \rangle &= \frac{V^2 \, | \langle \tilde{\delta}_1(\vec{k}) \, \tilde{\delta}_2^*(\vec{k}') \rangle |^2}{\tilde{Q}_1(0) \, \tilde{Q}_2(0)} \\
\langle \delta \hat{P}_1(\vec{k}) \, \delta \hat{P}_c(\vec{k}') \rangle &= \frac{V^2 \, {\rm Re} \left\{ \langle \tilde{\delta}_1(\vec{k}) \, \tilde{\delta}_1^*(\vec{k}') \rangle \, \langle \tilde{\delta}_1(\vec{k}) \, \tilde{\delta}_2^*(\vec{k'}) \rangle \right\}}{\tilde{Q}_1(0) \, \tilde{Q}_c(0)} \\
\langle \delta \hat{P}_c(\vec{k}) \, \delta \hat{P}_c(\vec{k}') \rangle &= \frac{V^2}{2 \, \tilde{Q}_c(0)^2} \, \{ |\langle \tilde{\delta}_1(\vec{k}) \, \tilde{\delta}_2^*(\vec{k}') \rangle|^2 + \nonumber \\
&{\rm Re} \left\{ \langle \tilde{\delta}_1(\vec{k}) \, \tilde{\delta}_1(\vec{k}') \rangle \langle \tilde{\delta}_2(\vec{k}) \, \tilde{\delta}_2(\vec{k}') \rangle \right\} \}
\end{align}
Writing $\delta \vec{k} = \vec{k} - \vec{k}'$ and using the
approximate relations in equations (\ref{eqd1d1}) and (\ref{eqd1d2})
gives
\begin{align}
&\langle \delta \hat{P}_1(\vec{k}) \, \delta \hat{P}_1(\vec{k}') \rangle = \frac{|P_1(\vec{k}) \, \tilde{Q}_1(\delta \vec{k}) + \tilde{S}_1(\delta \vec{k})|^2}{\tilde{Q}_1(0)^2} \\
&\langle \delta \hat{P}_1(\vec{k}) \, \delta \hat{P}_2(\vec{k}') \rangle = \frac{|P_c(\vec{k}) \, \tilde{Q}_c(\delta \vec{k})|^2}{\tilde{Q}_1(0) \, \tilde{Q}_2(0)} \\
&\langle \delta \hat{P}_1(\vec{k}) \, \delta \hat{P}_c(\vec{k}') \rangle = \nonumber \\
&\frac{{\rm Re} \left\{ \left[ P_1(\vec{k}) \, \tilde{Q}_1(\delta \vec{k}) + \tilde{S}_1(\delta \vec{k}) \right] \, P_c(\vec{k}) \, Q_c^*(\delta \vec{k}) \right\} }{\tilde{Q}_1(0) \, \tilde{Q}_c(0)} \\
&\langle \delta \hat{P}_c(\vec{k}) \, \delta \hat{P}_c(\vec{k}') \rangle = \frac{1}{2 \, \tilde{Q}_c(0)^2} \{ |P_c(\vec{k}) \, \tilde{Q}_c(\delta \vec{k})|^2 \nonumber \\
&+ {\rm Re} \left\{ \left[ P_1(\vec{k}) \, \tilde{Q}_1(\delta \vec{k}) + \tilde{S}_1(\delta \vec{k}) \right] \, \left[ P_2(\vec{k}) \, \tilde{Q}_2(\delta \vec{k}) + \tilde{S}_2(\delta \vec{k}) \right]^* \right\}
\end{align}
For a uniform selection function the equations simplify to:
\begin{align}
\langle \delta \hat{P}_1 \, \delta \hat{P}_1 \rangle &= \left( P_1 + \frac{1}{n_1} \right)^2 \\
\langle \delta \hat{P}_1 \, \delta \hat{P}_2 \rangle &= P_c^2 \\
\langle \delta \hat{P}_1 \, \delta \hat{P}_c \rangle &= P_c \left( P_1 + \frac{1}{n_1} \right) \\
\langle \delta \hat{P}_c \, \delta \hat{P}_c \rangle &= \frac{1}{2} \left[ P_c^2 + \left( P_1 + \frac{1}{n_1} \right) \left( P_2 + \frac{1}{n_2} \right) \right]
\end{align}
where $n_1 = N_1/V$ and $n_2 = N_2/V$.

\end{document}